\def\be{\begin{equation}}
\def\ee{\end{equation}}
\def\bea{\begin{eqnarray}}
\def\eea{\end{eqnarray}}
\def\sp{\;\;\;,\;\;\;}
\def\l{\lambda}
\def\lab{\label}
\def\f{\varphi}

\def\le{\left}
\def\ri{\right}

\def\half{\frac12}

\def\m{\mu}
\def\n{\nu}

\def\6{\partial}

\def\lab{\label}

\def\l{\lambda}

\def\tr{\textrm{tr}\,}
\def\m{\mu}
\def\n{\nu}
\def\bet{\begin{itemize}}
\def\eet{\end{itemize}}
\def\ben{\begin{enumerate}}
\def\een{\end{enumerate}}
\def\la{\langle}
\def\ra{\rangle}

\def\nn{\nonumber}

\documentclass[epj,nopacs]{svjour}
\usepackage{graphics}
\usepackage{graphicx}
\usepackage{amsmath,hyperref,amssymb}
\usepackage[usenames,dvipsnames,table]{xcolor}
\usepackage{comment,mathrsfs}
\usepackage{soul}
\usepackage{blindtext}
\usepackage{cuted}
\usepackage{flushend}
\begin{document}
\title{Holographic QCD and magnetic fields}
\author{Umut G\"ursoy%
}                     % Do not remove
\institute{Institute for Theoretical Physics, Utrecht University,
Leuvenlaan 4, 3584 CE Utrecht, The Netherlands}
\date{Received: date / Revised version: date}
\abstract{
We review the holographic approach to electromagnetic phenomena in large N QCD. After a brief discussion of earlier holographic models, we concentrate on the improved holographic QCD model extended to involve magnetically induced phenomena. We explore  the influence of magnetic fields on the QCD ground state, focusing on (inverse) magnetic catalysis of chiral condensate, investigate 
the phase diagram of the theory as a function of magnetic field, temperature and quark chemical potential, and, finally discuss effects of magnetic fields on the quark-anti-quark potential, shear viscosity, speed of sound and magnetization. 
\PACS{
      {PACS-key}{discribing text of that key}   \and
      {PACS-key}{discribing text of that key}
     } % end of PACS codes
} %end of abstract
\maketitle
%%%%%%%%%%%%%%%%%%%%%%%%%%%%%%%%%%%%%%%%%%%
%%%%%%%%%%%%%%%%%%%%%%%%%%%%%%%%%%%%%%%%%%%
%%%%%%%%%%%%%%%%%%%%%%%%%%%%%%%%%%%%%%%%%%%
\section{Introduction}
\label{sec::1}
%%%%%%%%%%%%%%%%%%%%%%%%%%%%%%%%%%%%%%%%%%%
%%%%%%%%%%%%%%%%%%%%%%%%%%%%%%%%%%%%%%%%%%%
%%%%%%%%%%%%%%%%%%%%%%%%%%%%%%%%%%%%%%%%%%%

Magnetic phenomena in QCD have been playing a major role in a variety of problems ranging from thermal properties of strongly interacting matter to its non-equilibrium evolution. A central motivation for a theoretical underpinning of electromagnetic properties of QCD is their realization in the ongoing and future planned experiments and observations. Intense magnetic fields are believed to occur in off-central heavy ion collisions \cite{Skokov:2009qp,Tuchin:2010vs,Voronyuk:2011jd,Deng:2012pc,Tuchin:2013ie,McLerran:2013hla,Gursoy:2014aka}. These fields, that are mostly produced by the spectator nuclei which do not participate in plasma formation and fly away from the collision region, can rise up to eB $\sim 5-10\  m_\pi^2$ where the pion mass $m_\pi$ provides a typical energy scale. As a result, these electromagnetic fields are expected to influence the charge dynamics in the quark-gluon plasma substantially. An example of such dynamics involves anomaly induced transport that is sourced by magnetic fields and/or vorticity \cite{Fukushima:2008xe,Kharzeev:2007jp,Tuchin:2010vs,Voronyuk:2011jd,Shi:2017cpu,Kharzeev:2015znc,Kharzeev:2010gd}. These, {\em chiral magnetic} and {\em chiral vortical} effects may have important implications for the strong CP problem and generation of baryon asymmetry in the early universe; see \cite{Kharzeev:2020jxw} for a recent review. It has been suggested that magnetic fields also induce more ordinary types of charge transport, arising from the Faraday effect and Lorentz and Coulomb forces in the quark-gluon plasma \cite{Gursoy:2014aka,Das:2016cwd,Gursoy:2018yai,Chatterjee:2018lsx} that would potentially leave distinctive imprints on experimental observables; see \cite{Dubla:2020bdz}. 

More recently, a fascinating new window into quark-gluon physics and its electromagnetic properties has opened by precision measurements of neutron stars \cite{TheLIGOScientific:2017qsa}, again highly magnetic and extremely dense astronomical objects that pack as much mass as two suns within a sphere of radius of about 10 km. The gravitational wave detectors LIGO and Virgo and the X-ray telescope NICER provide crucial new information on the Equation of State of quarks and gluons when they are dense and hot. Magnetic fields constitute an important part of the post-merger evolution and magneto-hydrodynamic description of QCD matter is essential in computing the post-merger gravitational wave and electromagnetic radiation profiles. With the very high rate of development in this field, it is conceivable that the gravitational waves and their electromagnetic counterparts \cite{GBM:2017lvd} arising from neutron star mergers and the subsequent kilonovae explosions would bring us new crucial information on the transport (electromagnetic and otherwise) properties of quark matter as well. 

All these developments point toward the need to underpin electromagnetic properties of QCD in a presumably strongly coupled regime. Electromagnetic phenomena in QCD is an age-old subject that has been studied using perturbative QCD, effective field theory \cite{Miransky:2015ava} and lattice QCD \cite{Bali:2011qj}; see \cite{Kharzeev:2012ph} for a review. These studies revealed interesting phenomena, such as the aforementioned {\em chiral magnetic effect}, {\em inverse magnetic catalysis} \cite{Bali:2012zg} and magnetically induced thermodynamic phases \cite{Miransky:2015ava}. However, if such electromagnetic effects arise when the QCD coupling is strong, then first-principles traditional methods --- with the exception of lattice QCD --- become hard to apply. On the other hand, lattice approach is mostly restricted to vanishing or small quark density and to static phenomena, not always suitable for studying neutron stars and transport in the QGP. Alternative non-perturbative methods, such as the functional renormalization group \cite{Pawlowski:2005xe}, and holographic correspondence \cite{Maldacena:1997re} are, therefore, welcome. This review  focuses on the use of holography to investigate electromagnetic properties of strongly interacting nuclear matter. We will refer to this relatively recent branch of research, perhaps too broadly, as {\em magneto-holography}.  

In the next section, after shortly introducing the basics of holography, we explain how to incorporate magnetic fields in this theory and review the relevant literature. In section \ref{sec::3} we focus on a specific model, called {\em improved holographic QCD}, and explain how to couple electromagnetic fields. In section \ref{sec::4} we explore the phase diagram of the theory at finite temperature, quark chemical potential and external magnetic field.  Section \ref{sec::5} is devoted to a discussion of thermodynamic observables such as the speed of sound and magnetization, and section \ref{sec::6} focuses on the physics of chiral condensate and how magnetic fields affect the ground state of large-N QCD. Holographic calculations reveal that the quark-anti-quark potential and hydrodynamic transport coefficients such as shear viscosity, depend strongly on the magnetic field as we explain in sections \ref{sec::7} and \ref{sec::8}. We end the review by discussing limitations of holography, providing a brief look at the topics we omitted, and, an outlook to open problems. The appendix provides details of the particular holographic model we consider. 
%%%%%%%%%%%%%%%%%%%%%%%%%%%%%%%%%%%%%%%%%%%
%%%%%%%%%%%%%%%%%%%%%%%%%%%%%%%%%%%%%%%%%%%
%%%%%%%%%%%%%%%%%%%%%%%%%%%%%%%%%%%%%%%%%%% 
\section{Magneto-holography}
\label{sec::2}
%1p
%%%%%%%%%%%%%%%%%%%%%%%%%%%%%%%%%%%%%%%%%%%
%%%%%%%%%%%%%%%%%%%%%%%%%%%%%%%%%%%%%%%%%%%
%%%%%%%%%%%%%%%%%%%%%%%%%%%%%%%%%%%%%%%%%%%
Gauge-gravity correspondence, ``holography'' for short,\\
stems from an equivalence between two separate descriptions of D-branes in terms of open and closed strings \cite{Maldacena:1997re,Gubser:1998bc,Witten:1998qj}. In the original example of Maldacena, this relates the maximally supersymmetric ${\cal N}=4$ super Yang-Mills theory to IIB closed string theory on the curved 10D $AdS_5 \times S^5$ space-time. However,  the idea of holography --- that is, the equivalence between gravitational theory in a $D$-dimensional space-time and gauge theory living on its $d<D$-dimensional boundary --- is believed to transcend 10D superstring theory and originate from more general principles \cite{Wilson:1974sk,tHooft:1993dmi,Susskind:1994vu}. Indeed, earlier examples of holography exist \cite{Kazakov:1985ea,Ginsparg:1993is,Polyakov:1998ju,Polyakov:2001af} and do not rely on D-branes. In this review, we follow the latter lore embodied in 5D {\em non-critical string theory} \cite{Polyakov:1998ju} and construct a 5D holographic theory for QCD in the presence of magnetic fields. 

%%%%%%%%%%%%%%%%%%%%%%%%%%
%%%%%%%%%%%%%%%%%%%%%%%%%%
\begin{figure}
\centering
\resizebox{.45\textwidth}{!}{
 \includegraphics{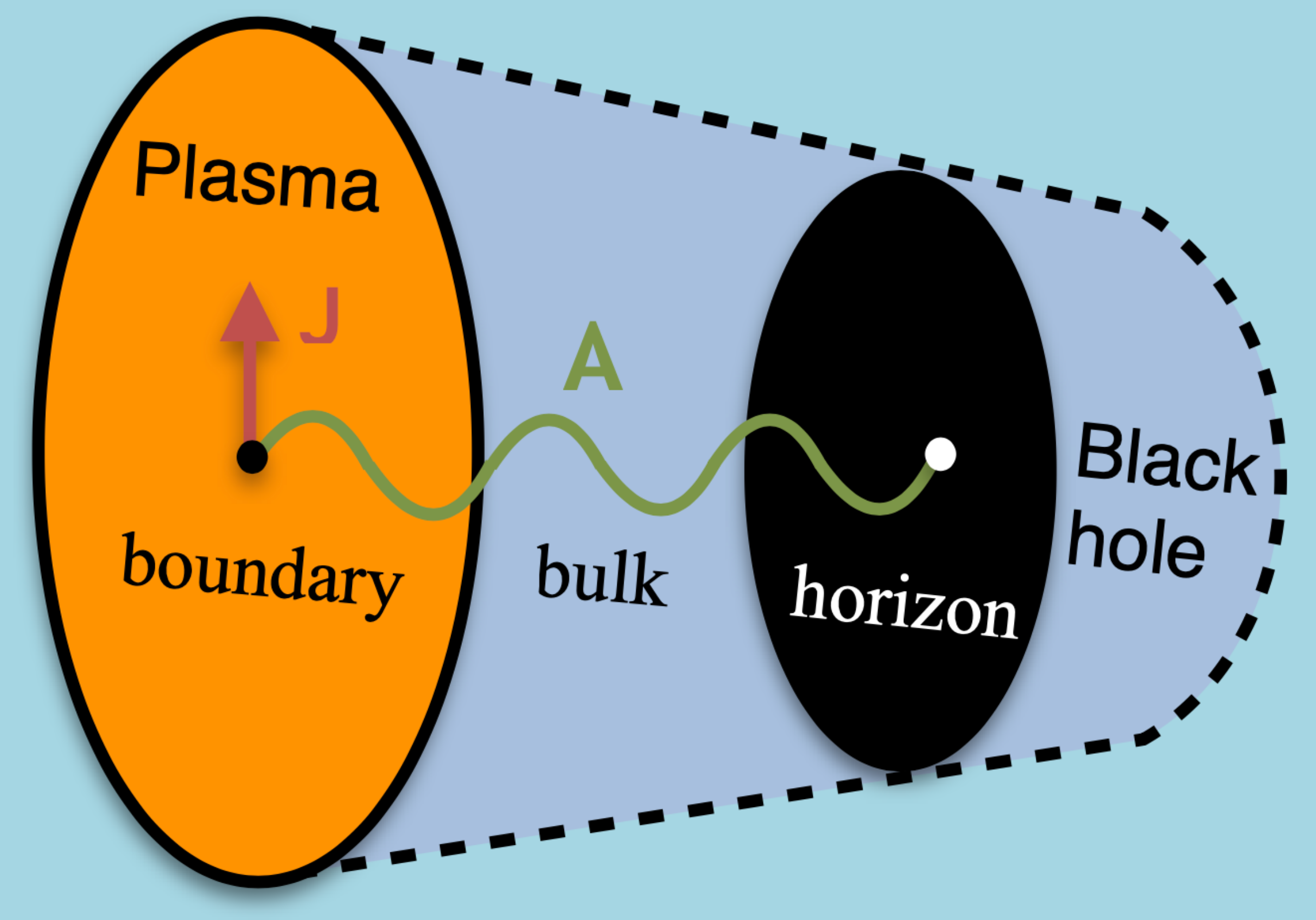}}
\vspace{.5cm}       
\caption{Holographic correspondence posits an equivalence between Einstein's general relativity in the bulk of a 5 dimensional hypothetical space with a black hole at the center and strongly coupled finite temperature quantum field theory at finite on the 4D boundary \cite{Maldacena:1997re,Gubser:1998bc,Witten:1998qj}. Concretely, it maps collective transport---here denoted by current J at the black point --- to fluctuations near the horizon --- at the white point---through propagation of bulk wave, A,  toward black hole. Universal properties of horizon geometry then lead to constraints on transport in the plasma.}
\label{fig::1}    
\end{figure}
%%%%%%%%%%%%%%%%%%%%%%%%%%
%%%%%%%%%%%%%%%%%%%%%%%%%%

Regardless of its origin, holography relates dynamics of closed strings in the bulk of an asymptotically AdS space-time to gauge theory that describes the dynamics of the boundary of this space. The additional non-compact {\em holographic coordinate} $r$ on the string side, corresponds to the renormalization group energy scale of the gauge theory where near boundary $r\to 0$ region corresponding to the UV. We will be interested in a thermal state on the boundary field theory, and this corresponds to having a black hole in the center of the bulk space \cite{Witten:1998zw}, see Fig.\ref{fig::1}. We will also consider the following Veneziano large N limit\cite{Veneziano:1979ec} in the gauge theory 
%%%%%%%%
\be\lab{veneziano}
N,\ N_f \to \infty, \quad g\to 0, \quad x\equiv \frac{N_f}{N},\quad \lambda \equiv g^2 N \gg 1\, ,
\ee 
%%%%%%%%
where $N$ is the number of colors, equivalently rank of the $SU(N)$ gauge group, $N_f$ is the number of quark flavors and $g$ is the Yang-Mills coupling constant. The combination $x$ and 't Hooft coupling $\lambda$ is kept fixed, but the latter is also taken large. This limit focuses on a simple, yet interesting corner of holography: Parameters of  gauge and string theory are related as \cite{Maldacena:1997re}
%%%%%%%%
\be\lab{dic1}
g_s \sim g^2, \qquad R\ell_s^2 \sim (g^2N)^{-\half} = \lambda^{-\half}\, ,
\ee
%%%%%%%%
where $g_s$ is the string coupling constant, $R\ell_s^2$ is the Ricci curvature of the gravitational background in string units and $\lambda$ is the 't Hooft coupling. The limit (\ref{veneziano}) then suppresses string loop corrections\footnote{Strictly speaking, it only suppresses closed string loops. Open string loops are also negligible in the gravitational backgrounds we work with \cite{Bigazzi:2005md}.} and the massive string states, reducing string theory to Einstein's gravity coupled to matter fields. The limit (\ref{veneziano}) is then motivated to have a simple, tractable corner of the holographic correspondence, and, at the same time, keeping intact the flavor sector to which electromagnetic fields couple. 

In this limit, the basic prescription of holography becomes equivalence between the generating function of the gauge theory and Einstein's action evaluated on-shell: 
%%%%%%%% 
\be\lab{GKPW}
{\cal W}[J(x)] = S_{gravity} [\phi(x, r_0) \to S(x)] \, ,
\ee
%%%%%%%%   
Here, ${\cal W}$ is the generating function of connected n-point functions, $S(x)$ is a source coupled to a gauge-invariant operator ${\cal O}(x)$, $\phi(x,r)$ is the corresponding bulk field --- see fig. \ref{fig::1} for an example where $A$ is  analogous to $\phi$ and $J$ is analogous to the VeV $\langle \cal O \rangle$ --- and $r_0$ is the boundary of the 5D geometry. Gravitational action in (\ref{GKPW}) includes a Gibbons-Hawking boundary term \cite{Gibbons:1976ue} to render metric variations well-defined, and counterterms \cite{Skenderis:2002wp} to remove divergences that arise when removing the cut-off $r_0\to 0$. The precise form of the limit in (\ref{GKPW}) as $r_0\to 0$ is \cite{Aharony:1999ti} 
%%%%%%%% 
\be\lab{bndry}
\phi(x, r) \to S(x) r^{4-\Delta} + R(x) r^\Delta\, ,\qquad r\to 0, 
\ee
%%%%%%%%
where $R$ is proportional to the VeV of the operator ${\cal O}$ and $\Delta$ is its conformal dimension in the UV theory. 

For example, RHS of (\ref{GKPW}) evaluated on a black hole with Hawking temperature $T$ and $S=0$, yields $F(T)/T$ where $F$ is the free energy of the gauge theory. This becomes relevant in section \ref{sec::5} where we investigate the thermodynamic properties of large N QCD. Connected correlators $\la {\cal O}(x) {\cal O}(y)\ra$ follow from evaluating the RHS, keeping $J$ as the boundary condition for $\phi$ and varying with respect to $S$ twice. The type of the correlator in Lorentzian time is determined by the other boundary condition in the deep interior of the 5D geometry. For example, retarded (advanced) Green's functions are given by the infalling (outgoing) boundary condition at the horizon \cite{Son:2002sd}. Retarded Green's function play an important role in transport --- see section \ref{sec::8} for an example --- as they determine hydrodynamic transport coefficients a la Kubo formulas \cite{Kubo:1957mj,Son:2007vk}.

%%%%%%%%%%%%%%%%%%%%%%%%%%%%%%%%%%%%%%%%%%%
%%%%%%%%%%%%%%%%%%%%%%%%%%%%%%%%%%%%%%%%%%%
%%%%%%%%%%%%%%%%%%%%%%%%%%%%%%%%%%%%%%%%%%% 
%\section{Magneto-holography}
%\label{sec::3}
%1p
%%%%%%%%%%%%%%%%%%%%%%%%%%%%%%%%%%%%%%%%%%%
%%%%%%%%%%%%%%%%%%%%%%%%%%%%%%%%%%%%%%%%%%%
%%%%%%%%%%%%%%%%%%%%%%%%%%%%%%%%%%%%%%%%%%%

Eventually, we are after the holographic description of SU(N) gauge theory coupled to U(1) electromagnetic gauge field. When U(1) interactions are weakly coupled, quantum fluctuations can safely be ignored and electromagnetism can be treated as a background gauge field coupled to a global current of the gauge theory. In ${\cal N}=4$ super Yang-Mills theory a proper proxy for this current is a U(1) subgroup of the $SU(4)$ R-symmetry of the theory \cite{Chamblin:1999tk}. In this theory, effect of $B$ on gauge dynamics is non-negligible without the need to scale the number of flavors as in (\ref{veneziano}), as the charged fields are all in adjoint representations that scale with the same power of $N$ as the gluons.

In the rest of this section we review the holographic description of ${\cal N}=4$ super Yang-Mills with an electromagnetic source. In section \ref{sec::3} we discuss a more realistic alternative with fundamental flavors charged under U(1) instead of adjoints, in the context of improved holographic models.

Generically, a background magnetic field $B$ in the $x_3$ direction in gauge theory corresponds to a dynamical {\em bulk} U(1) gauge field $A_\mu$ in the 5D bulk with the boundary condition
%%%%%%%%%%
\be
\lab{VmuB}
\lim_{r\to 0}A_\mu(r,x) = \left(\mu,-\frac{x_2 B}{2},\frac{x_1 B}{2},0,0\right)\, ,
\ee
%%%%%%%%%%
where we also added a time component, $\mu$, which will play the role of quark chemical potential. The holographic dual theory can be obtained by a twisted Kaluza-Klein reduction of IIB supergravity in 10D with the resulting action, 
%%%%%%%%%%
\be
\lab{act1}
S_{gravity} = \frac{1}{16\pi G_5} \int d^5x \sqrt{-g} \le(R + \frac{12}{\ell^2} -\ell^2 F^2 \ri) +CS  ,
\ee
%%%%%%%%%%
where $G_5$ is the 5D Newton's constant, $\ell$ is the AdS radius, $F=dA$ and there is a Chern-Simons term with a fixed coefficient. This is the bosonic part of minimal gauged supergravity in 5D \cite{Chamblin:1999tk}. The simplest solution to this action that is consistent with symmetries of a thermal state with non-vanishing $B$ and vanishing chemical potential  is of the form \cite{DHoker:2009mmn} 
%%%%%%%%%%
\bea
\lab{geo1}
ds^2 &=& e^{2A(r)} \le( \frac{dr^2}{f(r)} - dt^2 f(r) + e^{2W(r)}(dx_1^2 + dx_2^2) + dx_3^2 \ri), \nonumber \\  F &=& B dx_1\wedge dx_2\, ,
\eea
%%%%%%%%%%
where $f(r)$ is the blackening factor which vanishes at the horizon $f(r_h) = 0$. For the ground state with  $T\to 0$ one sets $f(r)=1$. The ground state preserves $SO(1,1) \times O(2)$ (boosts along and rotations around $B$) as in the dual field theory. The conformal factor $A(r)$ corresponds to the RG energy scale of the gauge theory\footnote{Intuitively this is because energy of a photon emitted from a point-like source at point $r$ in the interior should climb up the gravitational potential $\exp(A)$ before reaching the boundary implying that $A$ determines the energy of the source for the boundary observer.} \cite{Peet:1998wn} (see also \cite{deBoer:1999tgo,ihqcd1}). $W$ measures the anisotropy in the field theory caused by rotation symmetry breaking by $B$. 

Solution (\ref{geo1}) interpolates between $AdS_5$ near the boundary $r\to 0$ where $W\to 0$, $A\to -\log r$, $f\to 1$ and a BTZ black hole \cite{Banados:1992wn} times $R^2$ as $r\to r_h$. Analogously, the ground state exhibits an $AdS_3$ factor in the IR. This corresponds to the fact that an external magnetic field triggers an RG flow from the 4D conformal theory in the UV to an effectively 2D conformal theory in the IR as a result of Landau quantization \cite{DHoker:2012rlj}. Such $AdS_3$ factors will always be present in magneto-holography, albeit approximately, even in non-conformal holographic models more akin to QCD \cite{Gursoy:2020kjd}.

Before focusing on a specific magneto-holographic model, we list earlier holographic studies with magnetic fields\footnote{This is a representative selection with many unintended omissions.}:\cite{Erdmenger:2007bn,Albash:2007bq,Albash:2007bk,Johnson:2008vna,Bergman:2008sg,Bergman:2008qv,Preis:2010cq,Preis:2011sp,Preis:2012fh,Alam:2012fw,Rougemont:2014efa,Drwenski:2015sha,Rougemont:2015oea,Finazzo:2016mhm,Critelli:2016cvq,Gursoy:2017wzz}, see also \cite{Bergman:2012na} for an earlier review and the references therein. Phenomenon of (inverse) magnetic catalysis has been an important focus of previous holographic studies \cite{Filev:2007gb,Zayakin:2008cy,Filev:2011mt,Erdmenger:2011bw,Preis:2012fh,Mamo:2015dea,Dudal:2015wfn,Rougemont:2015oea,Evans:2016jzo,Jokela:2013qya,Gursoy:2016ofp}, as well as applications to condensed matter, see \cite{Hartnoll:2009sz} and references therein. Recently, there has also been revived interest in magnetic phenomena out of equilibrium in the context of heavy ion collisions \cite{Fuini:2015hba,Endrodi:2018ikq}.

%%%%%%%%%%%%%%%%%%%%%%%%%%%%%%%%%%%%%%%%%%%
%%%%%%%%%%%%%%%%%%%%%%%%%%%%%%%%%%%%%%%%%%%
%%%%%%%%%%%%%%%%%%%%%%%%%%%%%%%%%%%%%%%%%%%
\section{Improved holographic models}
\lab{sec::3}
%2p
%%%%%%%%%%%%%%%%%%%%%%%%%%%%%%%%%%%%%%%%%%%
%%%%%%%%%%%%%%%%%%%%%%%%%%%%%%%%%%%%%%%%%%%
%%%%%%%%%%%%%%%%%%%%%%%%%%%%%%%%%%%%%%%%%%%

In the rest of this review we concentrate on a specific holographic theory that shares the same salient features as large-N QCD: non-trivial RG flow with asymptotic freedom, linear confinement of quarks, gapped hadron spectrum, chiral symmetry breaking, a specific anomaly pattern and presence of a confinement/deconfinement phase transition at finite T. This {\em improved holographic QCD} theory \cite{ihqcd1,ihqcd2} follows the 5D non-critical string theoretic approach to holography and is constructed ``bottom-up''  i.e. by assuming that there exists a 5D string dual to QCD, approximating this string theory by Einstein's gravity coupled to matter fields, assumed to sufficiently capture the IR physics, and finally  constraining the Einstein's action by requiring the aforementioned properties of QCD. See \cite{ihqcd3} for a review of the improved holographic QCD program. See \cite{Ballon-Bayona:2021tzw} for a very recent and more comprehensive review of bottom-up holographic QCD.

%%%%%%%%%%%%%%%%%%%%%%%%%%
%%%%%%%%%%%%%%%%%%%%%%%%%%
\begin{figure}
\centering
\resizebox{.45\textwidth}{!}{
 \includegraphics{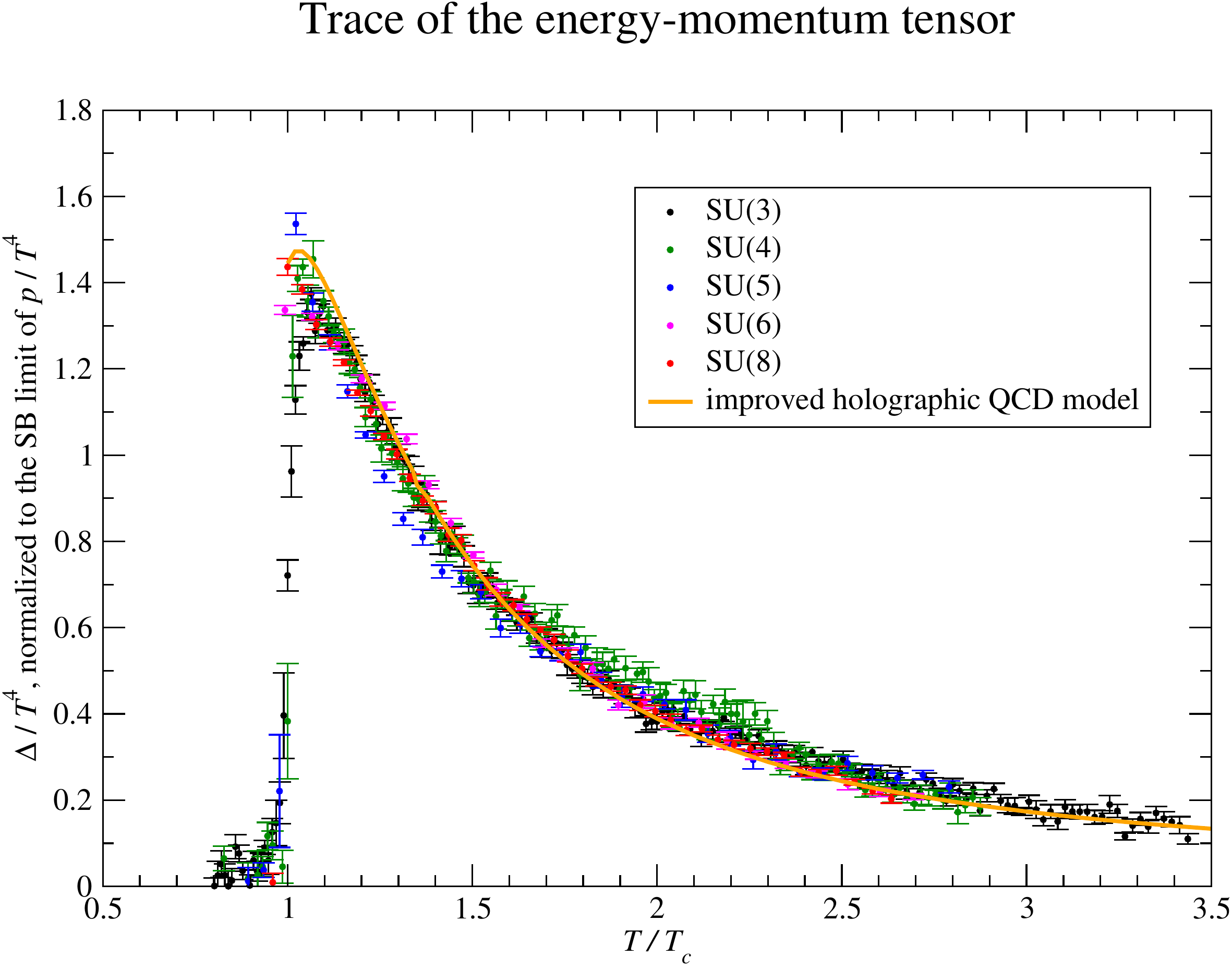}}
\vspace{.7cm}       
 \caption{Comparison of lattice QCD results to the improved holographic model for the trace of the stress tensor (interaction measure) normalized by $T^4 N^2$ in pure Yang-Mills theory. Lattice results are for $N=3,4,5,6$ and 8.}
 \lab{fig::n3}  
\end{figure}
%%%%%%%%%%%%%%%%%%%%%%%%%%
%%%%%%%%%%%%%%%%%%%%%%%%%%

Before detailing the holographic model, we present in Fig. \ref{fig::n3} a prediction of the holographic model for pure Yang-Mills, i.e. in the absence of quarks. We plot the so-called interaction measure --- i.e. the trace anomaly characterizing how far is the state from the conformal limit --- obtained from the holographic model and compared to lattice results. This figure demonstrates three points: (i) non-conformality of QCD/pure Yang-Mills is of utmost importance, especially around the deconfinement transition/cross-over which is the temperature range relevant for heavy ion collisions. (ii) when properly normalized by $N^2$, thermodynamic quantities become almost independent of $N$. This indirectly justifies our approach to QCD in the large N limit. (iii) holographic predictions agree very well with the lattice. Motivated by this agreement with the first-principles calculations when they are available, the rest of the paper reviews attempts to extend it with finite chemical potential and magnetic fields where lattice data becomes limited.   

%%%%%%%%%%%%%%%%%%%%%%%%%%
%%%%%%%%%%%%%%%%%%%%%%%%%%
\begin{figure}
\centering
\vspace{.4cm}
\resizebox{.5\textwidth}{!}{
 \includegraphics{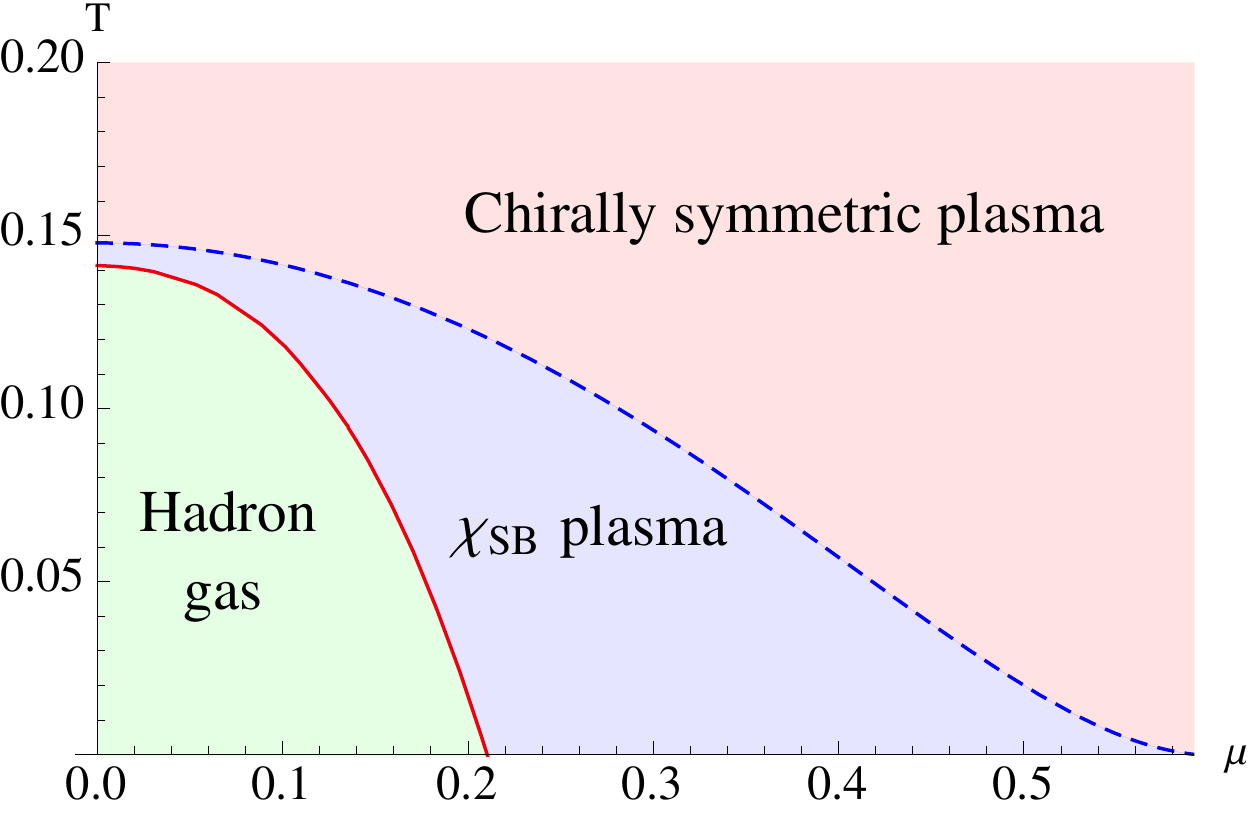}}
\vspace{.35cm}       
 \caption[]{Phase diagram of the ihQCD theory in the Veneziano limit with flavor-to-color ratio $x=1$, at finite quark chemical potential and vanishing magnetic field and quark masses. Axes are labelled in units of $\Lambda$. Figure is taken from \cite{Alho:2013hsa}.} 
\label{fig::2}    
\end{figure}
%%%%%%%%%%%%%%%%%%%%%%%%%%
%%%%%%%%%%%%%%%%%%%%%%%%%%

To determine which fields should be included in the 5D action one considers the bulk-boundary correspondence in (\ref{GKPW}) and introduces a 5D gravitational field per relevant and marginal operator in the IR: 5D metric dual to the stress tensor, a real scalar ``dilaton'' dual to $\tr G^2$ where $G_{\m\n}$ is the gluon field strength, a complex scalar ``tachyon'' dual to quark condensate, a 5D axion dual to $\tr G\wedge G$ and gauge fields dual to conserved global symmetries. Below, we will ignore the axion, as its effects are suppressed in the large N limit \cite{ihqcd2}, and consider only a single current $J$ dual to bulk U(1) gauge field (\ref{VmuB}). Our starting point is the following 5D gravity action that represent the glue and flavor contributions\footnote{The improved holographic model in the presence of fundamental flavors and in the Veneziano limit was coined ``V-QCD'' in \cite{Jarvinen:2011qe}. In what follows, we will use the term ``improved holographic QCD'' generally, to also include the V-QCD model.}, 
%%%%%%%%
\be\lab{fullaction} 
S_{gravity} = S_g + S_f\, ,
\ee 
%%%%%%%%
with the glue part $S_g$, given by 
%%%%%%%%
\be\lab{action} 
S_g = M_p^3 N^2 \int \sqrt{-g}~d^5 \le( R - \frac43 (\6\f)^2 + V(\f)\ri) + \textrm{GH} + S_{ct}\, ,
\ee
%%%%%%%%
where $R$ is the Einstein-Hilbert term, $\f$ is the dilaton. The potential $V$ in (\ref{action}) is assumed to include a negative cosmological constant, which, in the absence of $\f$, assures presence of an AdS solution. $M_p$ is the 5D Planck scale (a parameter of the model fixed by the UV limit of the free energy) and the $N$ dependence is factored out. The Gibbons-Hawking term, GH term is given by
%%%%%%%%
\begin{equation}
   {\cal S}_{GH} =2M_p^{3}N^2 \int_{\partial M}d^4x \sqrt{h}~K
    \label{app1}\end{equation}
%%%%%%%%
    with $K_{\m\n}$ denoting the extrinsic curvature
%%%%%%%%
\be
K_{\m\n}\equiv  -\nabla_\mu n_\nu = {1\over 2}n^{\rho}\partial_{\rho}h_{\m\n}\sp K=h^{ab}K_{ab}
\label{app2}
\ee
%%%%%%%%
and $h_{ab}$ is the induced metric on the boundary and $n_{\m}$ is the (outward) unit
vector normal to the boundary. 

%%%%%%%%%%%%%%%%%%%%%%%%%%
%%%%%%%%%%%%%%%%%%%%%%%%%%
\begin{figure*}[ht!]
\centering
\resizebox{.77\textwidth}{!}{
 \includegraphics{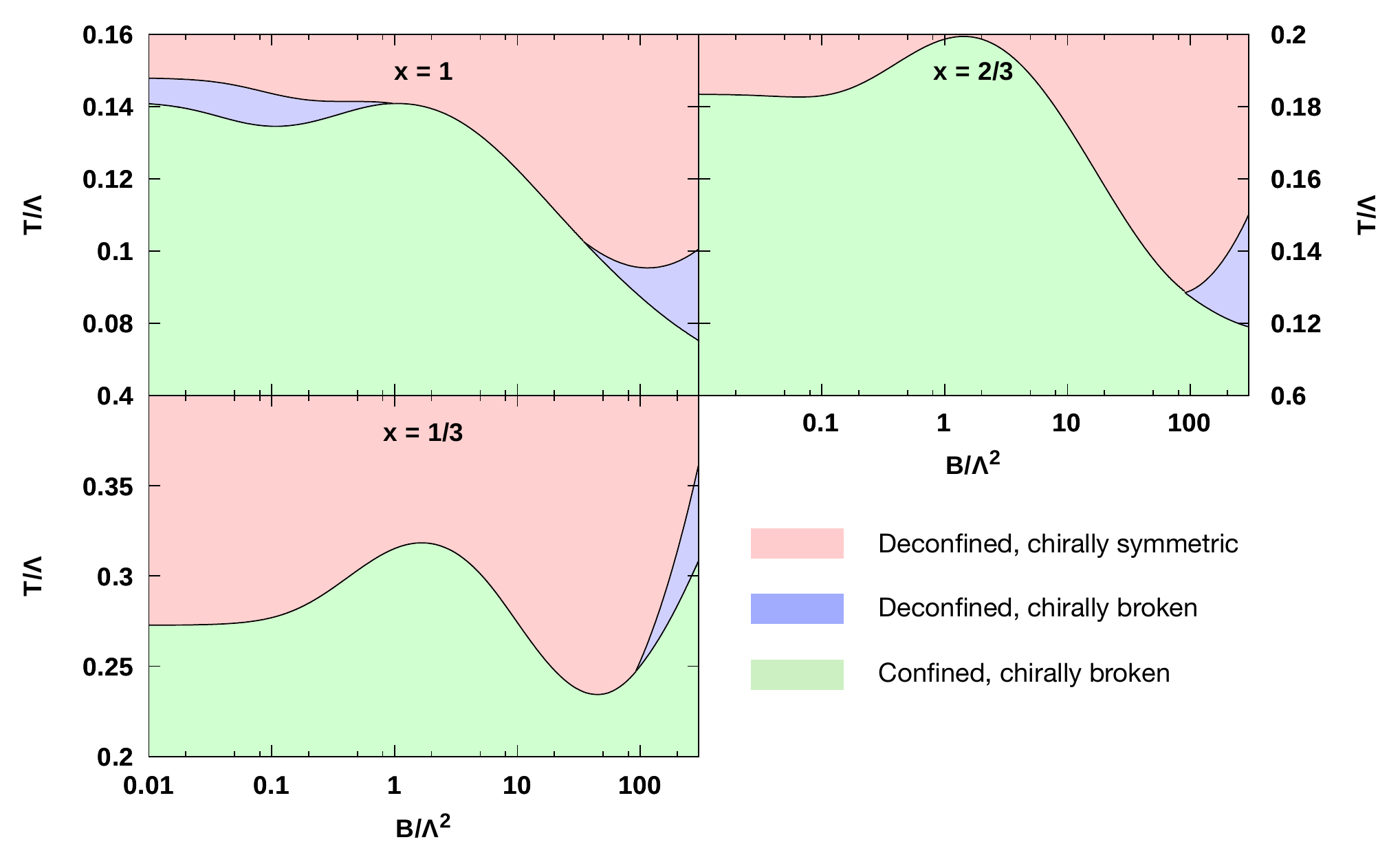}}
%\vspace{.5cm}       
 \caption{Phase diagram of ihQCD in the presence of an external magnetic field for various values of the ratio $x = N_f/N_c$. Figure is adapted from \cite{Gursoy:2016ofp}.}
\label{fig::3}    
\end{figure*}
%%%%%%%%%%%%%%%%%%%%%%%%%%
%%%%%%%%%%%%%%%%%%%%%%%%%%

The flavor part $S_f$ of (\ref{fullaction}) is given by the DBI action of $N_f+\overline{N}_f$ space-filling D4 flavor brane-anti-brane pairs for $N_f$ left and right handed quarks and their Wess-Zumino coupling to the background Ramond-Ramond fields\footnote{Wess-Zumino action includes a Chern-Simons term like in (\ref{act1}) and it is important for realization of QCD anomalies in holography\cite{Casero:2007ae}. It is also important to incorporate the contribution of dynamical gluons to anomalous transport \cite{Gursoy:2014ela}}. We will ignore the WZ term in this review for simplicity; see \cite{Arean:2016hcs,Gursoy:2014ela} for a full account of its relevance to anomalies and transport in QCD. The gauge theory living on the flavor branes correspond to the $U(N_f)_L \times U(N_f)_R$ flavor symmetry of the UV theory The $U(1)_{L-R}$ subgroup of this symmetry is absent in the quantum theory due to chiral anomaly. Below, we use the  $U(1)_{L+R}$ subgroup to introduce quark chemical potential and the magnetic field, as in (\ref{VmuB}). The generic form of this non-Abelian action is spelled out in \cite{Casero:2007ae,Jarvinen:2011qe}. We will make a further simplifying assumption and treat all quarks with the same mass and charge\footnote{With this assumption the flavor symmetry remains $SU(N_f)$ even in the presence of background magnetic field, instead of $SU(N_u)\times SU(N_d)$ for $N_u$ ``up-type'' quarks with charge +2/3 and $N_d$ ``down-type" quarks with charge -1/3.}. Then the flavor brane DBI action reduces to \cite{Jarvinen:2011qe}
%%%%%%%%
%\begin{strip}
%\begin{split}
\be 
\begin{aligned}
\label{actfs}
S_f &=-x\, M_p^3 N^2 \int d^5x\, V_f(\f,\tau) \sqrt{- \mathrm{det} D_{\m\n} } \, ,\\
D_{\m\n} &= g_{\m\n} + w(\f)\, F_{\m\n} + \kappa(\f)\, \partial_{\m} \tau \,\partial_{\n} \tau
\end{aligned}
\ee
%\end{split}
%\end{strip}
%%%%%%%%
where $x$ is the flavor to color ratio in the limit (\ref{veneziano}). We only turned on $U(1)_{L+R}$ subgroup of the full flavor symmetry with $F=dA$ the associated field strength, and set the other background gauge fields to zero. This gauge field satisfies the near boundary asymptotics (\ref{VmuB}). $\tau(r)$ denotes an open string ``tachyon'' dual to the quark mass operator with the near boundary asymptotics,
%%%%%%%%
\be
\tau(r) \simeq m_q r(-\log \Lambda r)^{- \rho}+\langle {\bar q} q \rangle r^3 (-\log \Lambda r)^{\rho}\, ,
\label{tauuv}
\ee
%%%%%%%%
as $r\to 0$. Here $\Lambda$ is a constant of motion proportional to $\Lambda_{QCD}$.  The power $\rho$ is to be matched to the anomalous dimension of ${\bar q} q$ and the QCD $\beta$-function (see~\cite{Jarvinen:2011qe,Arean:2013tja} for details). In this review we only consider massless quarks $m_q=0$, therefore the non normalizable mode of the tachyon solution vanishes, providing a boundary condition for the $\tau$ equation of motion. Then, the non-trivial profile of $\tau$ in $r$ corresponds to  spontaneous breaking of chiral symmetry in QCD. 

The general solutions with finite $B$ and $\mu$ will be of the form (\ref{geo1}), with addition of nontrivial temporal component of the gauge field --- with this, the U(1) field strength becomes $F = A_t'(r) dr \wedge dt + B dx_1 \wedge dx_2$ ---  together with nontrivial dilaton $\f$, and $\tau$ fields. This solution exhibits an IR singularity where the conformal factor vanishes as $A\to -\infty$ and the Einstein frame Ricci scalar diverges\footnote{Ricci scalar and other curvature invariants in the string frame $A_s = A + 2\f/3$ typically remain finite or vanish.}. A crucial condition for acceptability of this singularity is that it can be cloaked by an infinitesimal horizon and perturbative fluctuations around the background are repelled from the it \cite{Gubser:2000nd,ihqcd2}. The backgrounds in this review all satisfy these IR regularity criteria. 

%%%%%%%%%%%%%%%%%%%%%%%%%%
%%%%%%%%%%%%%%%%%%%%%%%%%%
\begin{figure*}[ht!]
\centering
\resizebox{.75\textwidth}{!}{
 \includegraphics{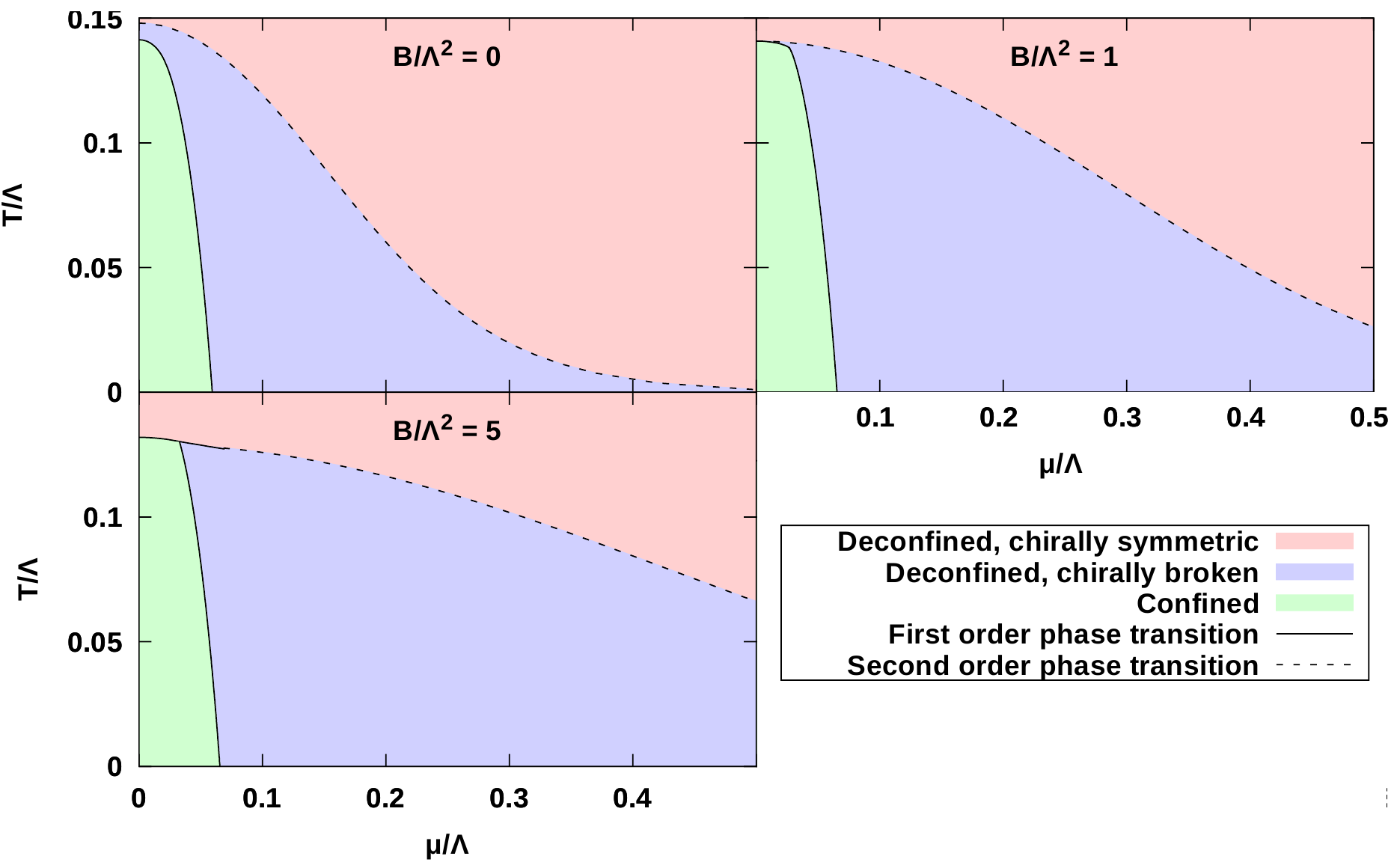}}
%\vspace{.6cm}       
 \caption[]{Phase diagram of the ihQCD theory in the Veneziano limit with flavor-to-color ratio $x=1$ and with finite quark chemical potential and for various magnetic fields. Figure from \cite{Gursoy:2017wzz}.} 
\label{fig::n1}    
\end{figure*}
%%%%%%%%%%%%%%%%%%%%%%%%%%
%%%%%%%%%%%%%%%%%%%%%%%%%%

The potentials $V$, $V_f$, $w$ and $\kappa$ should be chosen such that the resulting backgrounds satisfy basic properties of QCD. In particular, improved holographic QCD theory differs from other holographic constructions in the UV asymptotics of the dilaton potential $V$. $\exp(\f)$ couples to the gluon field strength $\tr G^2$, therefore corresponds to the `t Hooft coupling $\lambda$. On the other hand, as explained below (\ref{geo1}), the conformal factor $A$ is related to the RG energy scale as $A(r) = \log E$ in an holographic renormalization scheme \cite{ihqcd1}. Therefore the beta function of the field theory is determined from the ratio $\f'(r)/A'(r)$. Crucially, this ratio is in one-to-one connection to the dilaton potential $V$ upon using Einstein's equations and the IR regularity requirement above. All in all, this allows us fix the UV asymptotics of $V$ by reproducing asymptotic freedom in the UV. This leads to the small $\f$ asymptotics of $V$, 
%%%%%%%%
\be
\label{VUV}
V(\f) = \frac{12}{\ell^2} + v_1 e^\f + v_2 e^{2\f} + \cdots \, , \qquad \f\to-\infty\, ,
\ee
%%%%%%%%
where the first term is the cosmological constant and coefficients $v_i$ are fixed in terms of the beta-function coefficients of large N QCD\footnote{Demanding asymptotic freedom in holographic QCD should be understood as fixing the UV boundary conditions at a cut-off scale beyond which holographic description fails. Holographic duals of weakly coupled theories generally involve string corrections to all orders \cite{Gopakumar:2003ns}.}. On the other hand, the large $\f$ asymptotics of $V$ is fixed by the requirement of linear confinement of quarks \cite{ihqcd2} as, 
%%%%%%%%
\be
\label{VIR}
V(\f) = V_\infty\, e^{\frac43 \f} \f^\frac12 + \cdots \, , \qquad \f\to +\infty\, ,
\ee
%%%%%%%%
with $V_\infty$ some constant fixed by the quark string tension \cite{Gursoy:2010jh}. The choice of $V_f$  in (\ref{actfs}) was motivated in \cite{Bigazzi:2005md,Casero:2006pt,Casero:2007ae} to realize the spontaneous symmetry breaking and the axial anomaly of QCD, to be 
%%%%%%%%
\begin{equation}\lab{Vf}
V_f(\f,TT^{\dag})=V_{f0}(\f)e^{-a(\f)\tau^2}\, .
\end{equation}
%%%%%%%%
Similarly, UV and IR asymptotia of the other potentials entering (\ref{act1}) are determined by requiring other salient features of QCD \cite{Jarvinen:2011qe}. We list these potentials in appendix \ref{app::1} and refer to the literature for derivation, see e.g. \cite{Alho:2013hsa} and the references therein. One should stress that, even though large/small $\f$ asymptotia are fixed by the physical requirements, shape of the potentials for intermediate values of $\f$ remain undetermined to large extent. One typically introduces an efficient parametrization of these functions and fixes the remaining parameters by comparing with data from lattice-QCD or experiment. Presence of these systematic uncertainties is an inherent property of bottom-up holography which strongly affects quantitative predictions. 

%%%%%%%%%%%%%%%%%%%%%%%%%%
%%%%%%%%%%%%%%%%%%%%%%%%%%
\begin{figure*}[ht!]
\centering
\resizebox{.75\textwidth}{!}{
 \includegraphics{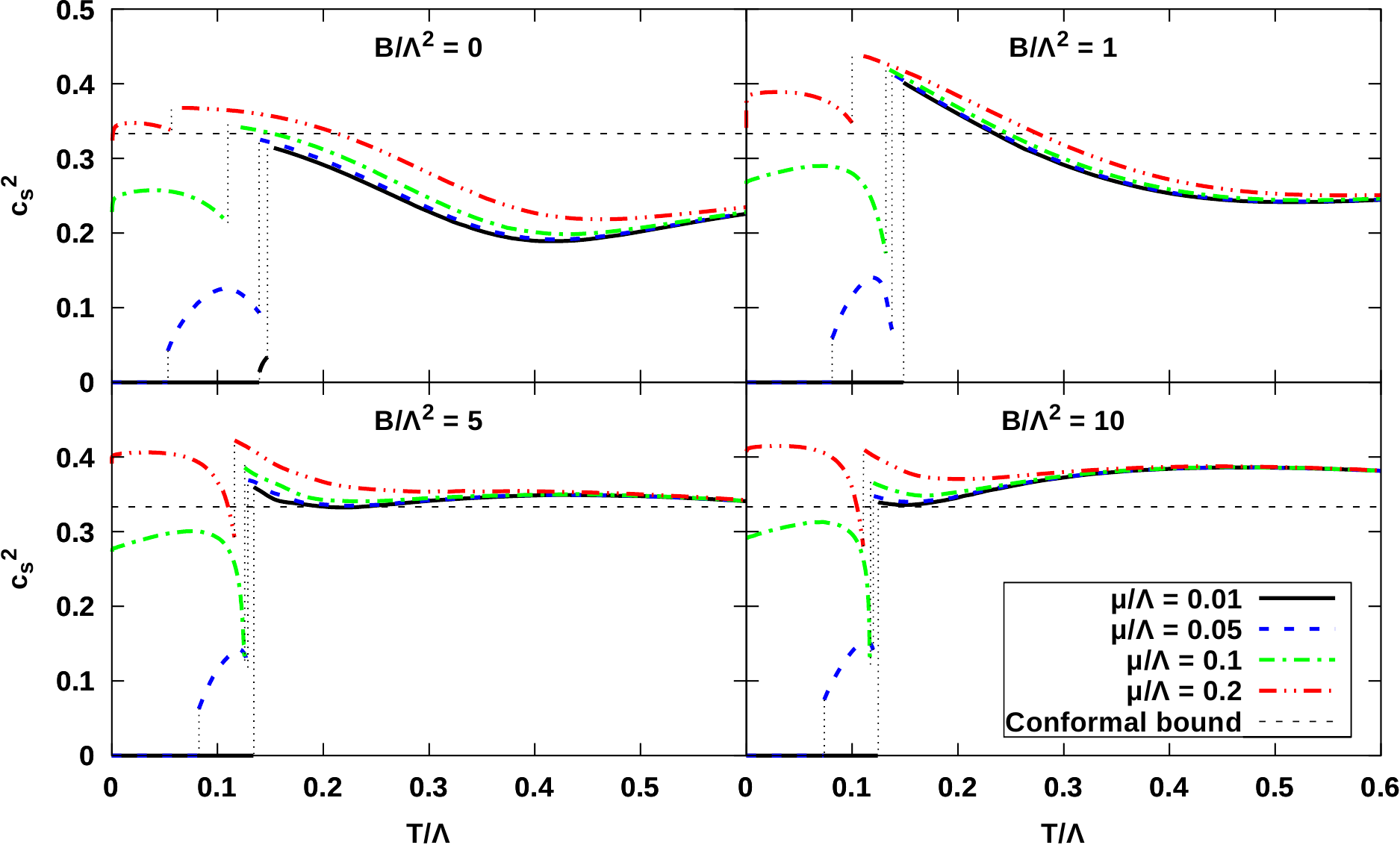}}
%\vspace{.5cm}       
 \caption{Speed of sound in holographic QCD as a function of temperature for different values of chemical potential and magnetic field. The dashed horizontal line corresponds to the conformal value $c_s^2 = 1/3$. Figure from \cite{Gursoy:2017wzz}.}
 \lab{fig::4}  
\end{figure*}
%%%%%%%%%%%%%%%%%%%%%%%%%%
%%%%%%%%%%%%%%%%%%%%%%%%%%

Qualitative predictions and understanding of universal features of dual gauge theories are unaffected by these systematic uncertainties, however. Some of the notable examples of such universal results are: the holographic c-theorem \cite{Freedman:1999gp}, bounds on charge and energy correlations \cite{Hofman:2008ar,Balakrishnan:2017bjg,Faulkner:2016mzt,Hartman:2016lgu} upper or lower bounds on transport coefficients \cite{Kovtun:2004de,Maldacena:2015waa,Grozdanov:2020koi}, one-to-one connection between existence of a deconfinement phase transition at $T_c$ and gapped and discrete hadron spectra \cite{ihqcd5}, existence of a maximum in bulk viscosity around $T_c$ \cite{Gursoy:2009kk},  $1/T^2$ scaling of the interaction measure $T^\mu_\mu/T^4$ for $T\gtrsim 1.5T_c$ \cite{Panero:2009wr,Caselle:2011mn}, bounds on speed of sound \cite{Hohler:2009tv,Cherman:2009tw}, modified Einstein relations between transport and diffusion \cite{Hatta:2008tx,Giecold:2009cg,Gursoy:2010aa}, universal relations between transport coefficients and thermodynamic properties \cite{Iqbal:2008by,Eling:2011ms} and anomalous transport coefficients at strong coupling \cite{Erdmenger:2008rm,Landsteiner:2012kd}. Holography was either directly used or was the inspiration behind these findings. Bottom-up holography is often the shortest path to such universal phenomena. 

%%%%%%%%%%%%%%%%%%%%%%%%%%%%%%%%%%%%%%%%%%%
%%%%%%%%%%%%%%%%%%%%%%%%%%%%%%%%%%%%%%%%%%%
%%%%%%%%%%%%%%%%%%%%%%%%%%%%%%%%%%%%%%%%%%%
\section{Phase diagram}
\lab{sec::4}
%2p
%%%%%%%%%%%%%%%%%%%%%%%%%%%%%%%%%%%%%%%%%%%
%%%%%%%%%%%%%%%%%%%%%%%%%%%%%%%%%%%%%%%%%%%
%%%%%%%%%%%%%%%%%%%%%%%%%%%%%%%%%%%%%%%%%%%

In this section, we summarize the qualitative picture arising from (\ref{actfs}), in particular how the phase diagram of the theory depends on quark chemical potential $\mu$ and external magnetic field $B$. To determine the phase diagram one obtains all asymptotically AdS solutions to (\ref{actfs}) with the boundary conditions as specified above, and compares their free energies in the grand canonical ensemble by evaluating corresponding on-shell values of the action, see equation (\ref{GKPW}). At vanishing $B$ and finite $\mu$, the holographic phase diagram was worked out in  \cite{Alho:2013hsa}, which we show in figure Fig. \ref{fig::2}. 

The confined phase, denoted by ``hadron gas'' in the figure, exists up to some finite $\mu$ in the small temperature regime. In this phase the chiral symmetry is spontaneously broken as $SU(N_f)_L\times SU(N_f)_R\to SU(N_f)_{L+R}$ by non-vanishing chiral condensate. This phase is represented by the ``thermal gas solution'' in the holographic background. It is a horizonless solution obtained from (\ref{geo1}) by substituting $f=1$ and $W=0$. The phase denoted by $\chi_{SB}$  is a deconfined quark-gluon plasma with non-vanishing chiral condensate; i.e.  chiral symmetry remains broken. Holographically, this phase corresponds to a black hole accompanied by a non-trivial vector bulk field and a non-trivial profile for the tachyon field $\tau(r)$. The hadron gas phase is separated from the $\chi_{SB}$ phase by a first order phase separation curve $T_c(\mu)$ (red, solid)  in figure \ref{fig::2}. At higher temperatures, the chiral condensate melts through a second-order phase transition (blue, dashed curve) at $T_\chi(\mu)$ resulting in a deconfined state with chiral symmetry restoration. This phase boundary becomes a continuous crossover when quark masses are included \cite{Alho:2013hsa}. This high temperature (pink in Fig.) phase holographically corresponds to  a black hole  with $\tau=0$. 

Generic features of the phase diagram remains the same for different choices of the potentials in the action, in particular the existence of the three phases and the nature of the phase boundaries remain unaltered. Precise location of the phase boundaries will depend on the details of the bottom-up construction. Unfortunately, not much can be said about the location of the critical point conjectured to be present on the $(T,\mu)$ plane in real QCD \cite{Stephanov:1999zu}, as this critical point is pushed to the $\mu=0$ boundary of the phase diagram in the Veneziano limit, the limit we employ in this paper. 

%%%%%%%%%%%%%%%%%%%%%%%%%%
%%%%%%%%%%%%%%%%%%%%%%%%%%
\begin{figure*}[ht!]
\centering
\resizebox{.77\textwidth}{!}{
 \includegraphics{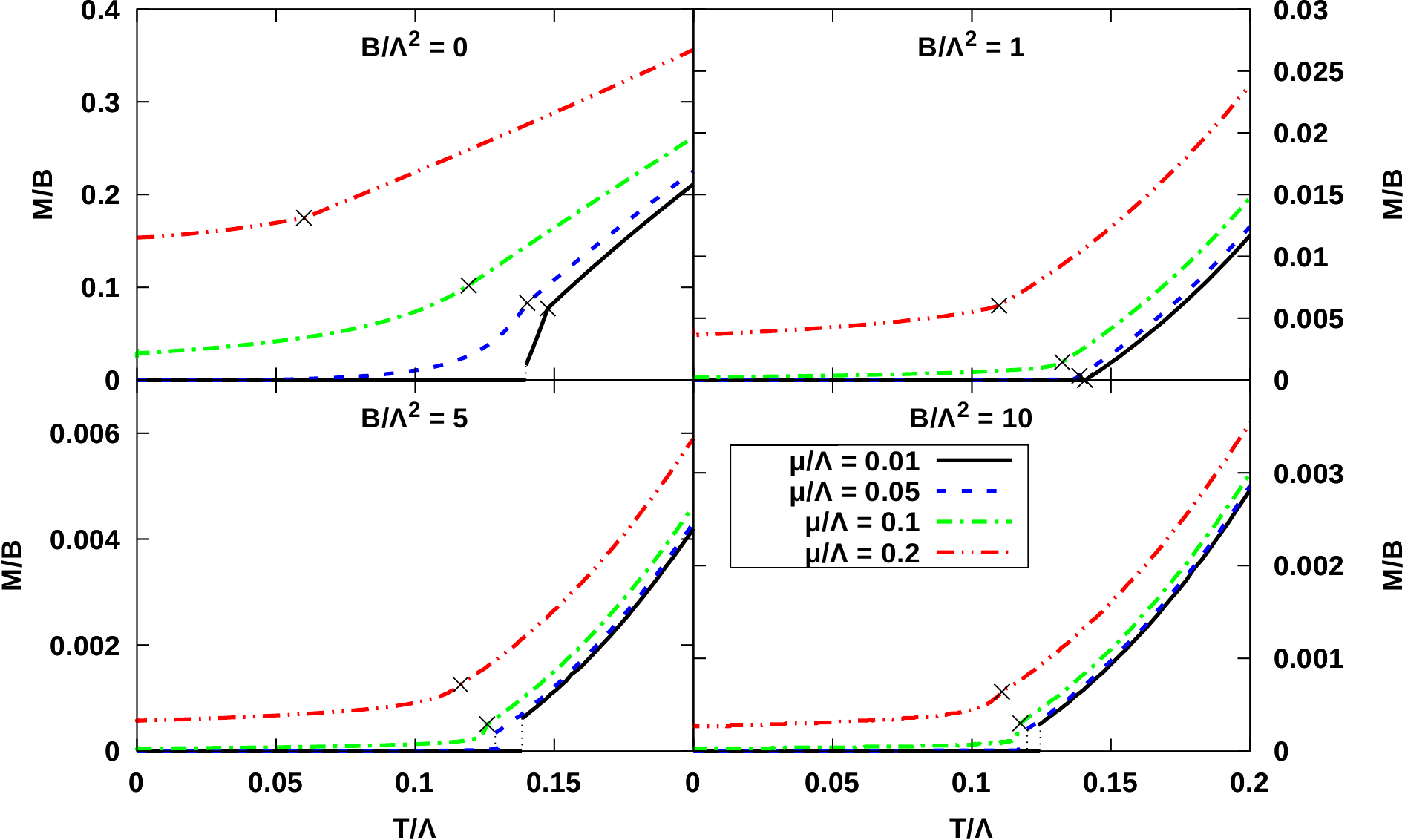}}
%\vspace{.5cm}       
 \caption{Magnetization normalized by the magnetic field strength as a function of temperature for various choices of $\mu$ and $B$.  The crosses denote chiral symmetry restoration transitions. M
is normalized such that $M(T=0,\mu=0) = 0$ for all $B$. Figure from \cite{Gursoy:2016ofp}.}
 \lab{fig::n2}  
\end{figure*}
%%%%%%%%%%%%%%%%%%%%%%%%%%
%%%%%%%%%%%%%%%%%%%%%%%%%%

Phase diagram at vanishing $\mu$, finite $B$ in the improved holographic model was studied in \cite{Gursoy:2016ofp} and the result is shown in Fig. \ref{fig::3}. In this figure we also show how the flavor-to-color ratio $x$ influences the diagram. We observe that presence of the intermediate (blue) phase  depends on the ratio $x$ and the magnitude of $B$. In particular, it disappears for smaller $x$ at small values of $B$. It reappears at higher magnetic fields for all $x$. For sufficiently large $x$, the phase transition temperatures between the hadronic and the plasma phase and the chirally broken and unbroken plasmas decrease with increasing $B$ for small $B$. As we discuss in the next section, this will be connected to the inverse magnetic catalysis phenomenon.  A 3D holographic phase diagram including all axes $(T,\mu,B)$ is not available yet. A priority in this regard is exploration of the other conjectured phases in QCD such as baryonic, color-flavor-locked and superconducting phases. These are all interesting future directions that require solutions to conceptual issues such as how to realize baryons in bottom-up holographic QCD \cite{Ishii:2019gta,Giataganas:2018uuw}. 

Finally, in Fig. \ref{fig::n1} we present the phase diagram on the temperature-chemical potential plane for the various choices of $B$. We included $B=0$ case (same as Fig. \ref{fig::2}) on the top left corner as a reference point. Even though magnetic field tends to remove the chirally broken plasma (blue) phase at $\mu=0$, this phase is present at finite $\mu$ and $B$, and remains a generic prediction of the holographic model. Whether there is an analog in actual QCD is unclear, it may well be cloaked by baryonic, superconducting or color-flavor locked phases, which have not yet been fully investigated as possible saddles of the dual gravitational theory. Another feature emerging in Fig. \ref{fig::n1} is the merger of first order confinement/deconfinement transition with the second order chiral symmetry restoration transition at finite $\mu$ and sufficiently large $B$. Close inspection of the merging point reveals a bifurcation of the solid and the dashed lines and the solid line ending on a critical point; see \cite{Gursoy:2017wzz} for details. 

%%%%%%%%%%%%%%%%%%%%%%%%%%%%%%%%%%%%%%%%%%%
%%%%%%%%%%%%%%%%%%%%%%%%%%%%%%%%%%%%%%%%%%%
%%%%%%%%%%%%%%%%%%%%%%%%%%%%%%%%%%%%%%%%%%%
\section{Thermodynamic observables}
\lab{sec::5}
%2p
%%%%%%%%%%%%%%%%%%%%%%%%%%%%%%%%%%%%%%%%%%%
%%%%%%%%%%%%%%%%%%%%%%%%%%%%%%%%%%%%%%%%%%%
%%%%%%%%%%%%%%%%%%%%%%%%%%%%%%%%%%%%%%%%%%%

An essential quantity that characterizes the thermodynamic equation of state in the deconfined phase is the {\em speed of sound} $c_s$. Sound waves are comprised of alternating high and low pressure regions along the direction of the wave motion. In the absence of magnetic field and chemical potential, the typical speed of sound waves are determined by how pressure density varies with the energy density, 
%%%%%%%%%%
\be\lab{cs2} 
c_s^2 =  \frac{dp}{d\epsilon} = \frac{s dT}{T ds} \, ,
\ee
%%%%%%%%%%
where we used the first law of thermodynamics in the second equation. At finite $\mu$ and $B$, this is generalized to 
%%%%%%%%%%
\be\lab{cs2muB} 
c_s^2 =  \frac{s dT + n d\mu}{T ds + \mu dn + B dM}\bigg|_{n/s,B} \, ,
\ee
%%%%%%%%%%
where $n$ is quark density and $M$ is magnetization. Constraints on $c_s$ arising from microscopic physics would be extremely interesting both for the QGP and for the neutron stars \cite{Annala:2017llu}. Such constraints can be derived from holography at strong coupling within certain assumptions. Refs. \cite{Hohler:2009tv,Cherman:2009tw} showed that $c_s$ approaches to its conformal value $1/\sqrt{3}$ {\em from below} at high temperatures when conformality is broken by a single relevant scalar operator. The latter is also the situation in improved holographic QCD in the absence of flavors, that is, in the limit of large N pure Yang-Mills theory. The question remains whether similar holographic bounds arise at finite $\mu$ and/or $B$. This was studied in \cite{Gursoy:2017wzz} and the outcome is shown in Figs. \ref{fig::4}. As clear from these figures that the proposed conformal bound is violated at finite chemical potential and non-vanishing magnetic field. There is of course no contradiction as \cite{Hohler:2009tv,Cherman:2009tw} assumed $\mu=B=0$.  Whether, there is another universal upper bound (smaller than the speed of light) at strong coupling, also at finite $\mu$ and $B$ remains to be seen.

Another observable of interest is {\em magnetization} $M$. This is defined as the response of energy (or free energy) to variation of the external magnetic field
%%%%%%%%%%
\be\lab{mag} 
M =  -\frac{dF}{dB}\bigg|_{T,\mu} \, ,
\ee
%%%%%%%%%%
and it has been studied at vanishing quark chemical potential by lattice QCD simulations, see e.g \cite{Bali:2014kia}.  We plot magnetization in Fig. \ref{fig::n2} as a function of $T$ for the various values of $B$ and $\mu$. Magnetic susceptibility which is defined as the derivative of $M$ with respect to $B$ is also computed in \cite{Gursoy:2016ofp} for vanishing $\mu$. In the limit $B\to 0$ it can be read off from  from the top-left corner of  Fig. \ref{fig::n2} as it is equivalent to $M/B$ in the limit $B=0$. 

Magnetization is directly linked to the shape of phase boundaries in the phase diagram with $B$. In particular, it determines how the deconfinement (solid curve in Fig. \ref{fig::n1}) and the chiral symmetry restoration (dashed curve) transition temperatures vary with the magnetic field \cite{Ballon-Bayona:2017dvv,Gursoy:2017wzz}. If one holds fixed the chemical potential, then it immediately follows from the first law $dF = - S dT - M dB$ and the fact that the free energy is continuous across a phase transition that deconfinement temperature satisfies 
%%%%%%%%%%
\be\lab{Td} 
\frac{\delta T_d(B)}{\delta B}  = - \frac{\Delta M(B)}{\Delta S(B)}  \, ,
\ee
%%%%%%%%%%
where $\Delta M$ and $\Delta S$ are difference of the quantity across the phase boundary. In a deconfinement phase transition $\Delta S > 0$ hence the sign of the LHS is completely determined by $\Delta M$. Applying the same reasoning to the second-order chiral restoration transition (with $\Delta S =0$) one finds that $\Delta M = 0$ and that the chiral restoration temperature satisfies 
%%%%%%%%%%
\be\lab{Tchi} 
\frac{\delta T_\chi(B)}{\delta B}  = - \frac{\Delta \frac{\6M}{\6T}}{\Delta \frac{\6S}{\6 T}}  \, .
\ee
%%%%%%%%%%
Noting that the denominator is proportional to the difference of specific heats across the phase boundary, one concludes that $T_\chi$ increases (decreases) with $B$ iff $\6 M/ \6T $ in the chiral symmetric phase is higher (lower) than the same quantity in the chirally broken phase. This will become relevant when assessing (inverse) magnetic catalysis of the chiral condensate in the next section. 

%%%%%%%%%%%%%%%%%%%%%%%%%%%%%%%%%%%%%%%%%%%
%%%%%%%%%%%%%%%%%%%%%%%%%%%%%%%%%%%%%%%%%%%
%%%%%%%%%%%%%%%%%%%%%%%%%%%%%%%%%%%%%%%%%%%
\section{Ground state and inverse magnetic catalysis}
\lab{sec::6}
%2p
%%%%%%%%%%%%%%%%%%%%%%%%%%%%%%%%%%%%%%%%%%%
%%%%%%%%%%%%%%%%%%%%%%%%%%%%%%%%%%%%%%%%%%%
%%%%%%%%%%%%%%%%%%%%%%%%%%%%%%%%%%%%%%%%%%%

Perturbative QCD and effective field theory studies generically show that the chiral condensate is strengthened in the presence of a magnetic field; a phenomenon called ``magnetic catalysis'' \cite{Gusynin:1994re,Gusynin:1994xp,Gusynin:1994va}. This can be qualitatively understood as an outcome of Landau quantization: motion of quarks transverse to $B$ are restricted resulting in reduction from 3+1 to 1+1 in the effective dimension of the system. As the IR effects responsible for condensate formation are stronger in 2D, this results in an effective increase in the magnitude of the condensate with $B$. This suggestive argument is substantiated by explicit calculations at weak coupling, see \cite{Miransky:2015ava} for a review.  

%%%%%%%%%%%%%%%%%%%%%%%%%%
%%%%%%%%%%%%%%%%%%%%%%%%%%
\begin{figure}
\centering
\resizebox{.45\textwidth}{!}{
 \includegraphics{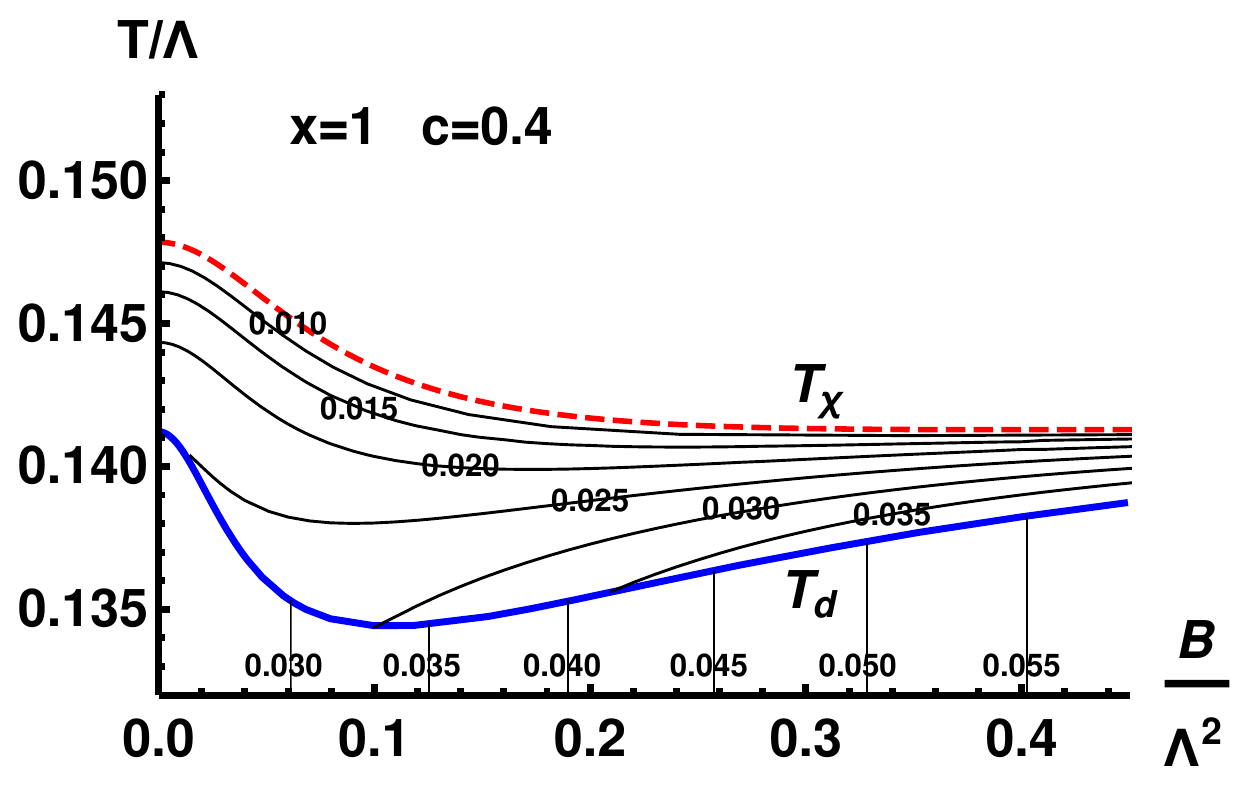}}
\vspace{.5cm}       
 \caption{Phase boundaries and curves of constant $\la \bar q q\ra$ in ihQCD for a choice of $x=1$ and $c=0.4$ (a parametrization of the holographic potential $w$, see the appendix).  Dimensionful quantities are normalized by $\Lambda\approx 1 GeV$. Plot reproduced from paper \cite{Gursoy:2016ofp}.}
\label{fig::5}    
\end{figure}
%%%%%%%%%%%%%%%%%%%%%%%%%%
%%%%%%%%%%%%%%%%%%%%%%%%%%

However, lattice studies with 2+1 flavors \cite{Bali:2011qj,Bali:2011uf,Bali:2012zg,D'Elia:2012tr} revealed a more complicated behavior. It is found that for temperatures smaller than the deconfinement transition temperature---that is around  $150$~MeV---the condensate in the confined phase increases with $B$ up to a certain turning point, above which it decreases with increasing $B$. This turning point depends on the temperature. Moreover, for larger temperatures slightly below deconfinement, the condensate decreases even for small B. Therefore one finds that strong coupling in QCD triggers the effect opposite to magnetic catalysis. This is coined ``inverse magnetic catalysis'' (IMC). 

The precise physical mechanism for this behavior is not completely clear. There are indications from further lattice studies \cite{Bruckmann:2013oba,Bruckmann:2013ufa} that the presence of a turning point in the condensate as a function of $B$ results from a competition between two separate contributions. Considering the path integral representation of $\la \bar q q \ra$, one can identify these two contributions as follows. 1) a direct coupling of fermion propagators inside $\bar q q$  to B, called  ``valence quarks'' in \cite{Bruckmann:2013oba}. This always tends to strengthen the condensate, essentially for the same reason as above that leads to magnetic catalysis. 2) an indirect coupling of B to the quark determinant arising from the gluon path integral, called ``sea quarks''. This contribution becomes stronger at  intermediate  or large values of the coupling constant, and it was argued in \cite{Bruckmann:2013oba,Bruckmann:2013ufa} that it dominates over the first source for relatively large values of B and T, leading to the inverse effect.  See \cite{Mueller:2015fka} for a similar suggestion where the authors propose that IMC results from a combined effect of gluon screening and weakening of gauge coupling at high energies. These are, however, mostly suggestive arguments and it is desirable to investigate the question using an alternative non-perturbative approach.

%%%%%%%%%%%%%%%%%%%%%%%%%%
%%%%%%%%%%%%%%%%%%%%%%%%%%
\begin{figure}
\centering
\resizebox{.45\textwidth}{!}{
 \includegraphics{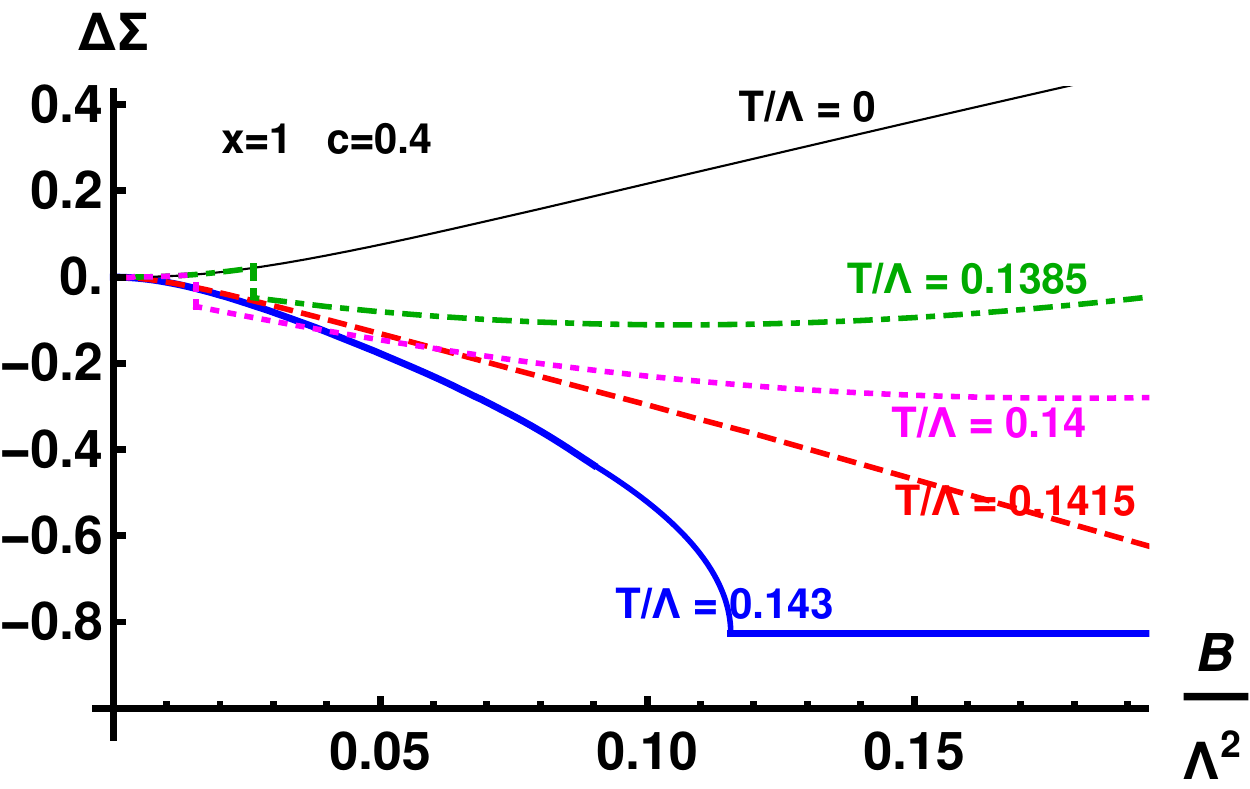}}
\vspace{.5cm}       
 \caption{(Normalized) chiral condensate as a function of B in the deconfined - chirally broken phase in ihQCD for $x=1$ and $c=0.4$. This picture demonstrates the phenomenon of inverse magnetic catalysis.  Plot reproduced from paper \cite{Gursoy:2016ofp}.}
\label{fig::6}    
\end{figure}
%%%%%%%%%%%%%%%%%%%%%%%%%%
%%%%%%%%%%%%%%%%%%%%%%%%%%

The question has been addressed in holography in the various works \cite{Zayakin:2008cy,Filev:2011mt,Erdmenger:2011bw,Preis:2012fh,Mamo:2015dea,Dudal:2015wfn,Rougemont:2015oea,Evans:2016jzo,Jokela:2013qya} for toy systems involving adjoint flavors or small number of fundamental quarks for which the fermion contribution to the background is suppressed at large N. Recently, the question has been investigated in detail in \cite{Gursoy:2016ofp} in the Veneziano limit (\ref{veneziano}) where it was concluded that the holographic description supports the valence vs. sea quark suggestion of \cite{Bruckmann:2013oba,Bruckmann:2013ufa}. 

In the rest of this section, we present the findings of \cite{Gursoy:2016ofp}. Effect of B on the ground state and the phase diagram can be parametrized by a constant $c$ that enters the $w$ potential in (\ref{actfs}), see Appendix. We find that inverse magnetic catalysis takes place for small choices of this constant, which we will choose as either\footnote{It becomes clear below that allowing a small freedom in this choice is beneficial in gauging the effects of magnetic fields in holography. Fine-tuning of this constant requires a more extensive matching of holographic QCD with lattice.} of $c=0.4$ or $c=0.25$.  In figure \ref{fig::5} we show the phase boundaries for the deconfinement and chiral symmetry restoration transitions for $x=1$ and $c=0.4$. We observe that both of these transition temperatures initially decrease with increasing $B$. In the deconfined - chiral symmetry broken phase with $T_d<T<T_\chi$, this means that it becomes easier to melt the condensate at larger $B$. Contours of constant condensate are given in the same plot. A look at these contours in the phase $T_d<T<T_\chi$ confirms that the condensate decreases with B for sufficiently small values, as expected from IMC. One also observes that the curves of constant condensate extend between the curves $T_d(B)$ and $T_\chi(B)$ continuously decreasing with increasing T and finally vanishing at $T_\chi$ leading to the second order chiral symmetry restoration transition discussed in the previous section\footnote{Constancy of the condensate (see vertical contours in Fig. \ref{fig::5}) below $T_d$ is an artifact of holographic QCD as the temperature dependence in the confined phase is suppressed in the large-N limit.}.   

%%%%%%%%%%%%%%%%%%%%%%%%%%
%%%%%%%%%%%%%%%%%%%%%%%%%%
%\begin{strip}
\begin{figure}
\centering
\resizebox{.45\textwidth}{!}{
 \includegraphics{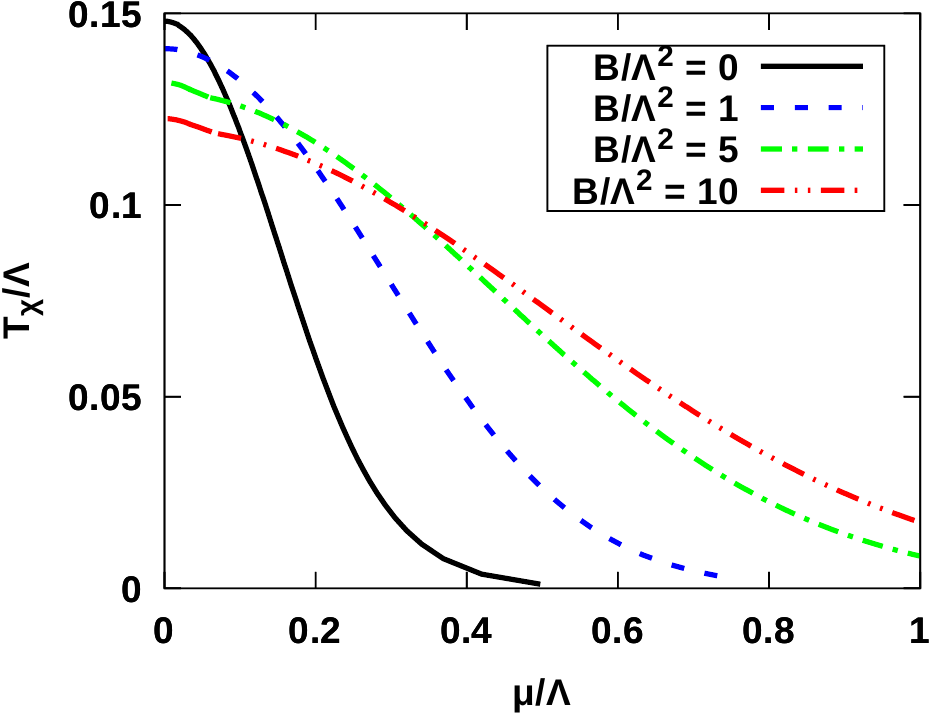}}
\vspace{.3cm}       
 \caption{Chiral symmetry restoration temperature as a function of quark chemical potential for various values of B. Figure from \cite{Gursoy:2017wzz}.}
\label{fig::7}    
\end{figure}
%\end{strip}
%%%%%%%%%%%%%%%%%%%%%%%%%%
%%%%%%%%%%%%%%%%%%%%%%%%%%

Finally, in figure \ref{fig::6} we plot the renormalization invariant and dimensionless combination $\Delta \Sigma(T,B) =\Sigma(T,B)-\Sigma(T,0)$ where 
 %%%%%%%%%%
\be
\Sigma(T,B)= \frac{\langle \bar q q \rangle (T,B)}{\langle \bar q q \rangle (0,0)}\, .
\ee
 %%%%%%%%%%
We observe, in qualitative agreement with the lattice results \cite{Bali:2011qj,Bali:2011uf,Bali:2012zg,D'Elia:2012tr} mentioned above, that the condensate increases with B up to a certain temperature around $T/\Lambda\approx 0.138$, then it starts decreasing for larger T until restoration of chiral symmetry. For larger $T$ the condensate drops to zero for large $B$, as demonstrated by the blue curve in figure \ref{fig::6}, because the condensates vanishes above $T_\chi$, cf. fig. \ref{fig::5}. 

A suggestive argument for the physical mechanism behind inverse magnetic catalysis, similar to the argument of  \cite{Bruckmann:2013oba,Bruckmann:2013ufa}---i.e. competition between ``valence'' vs.  ``sea''---can also be made in holography. The condensate is determined by the equation of motion of $\tau$, see (\ref{actfs}). This equation depends on $B$ again in two separate ways: an explicit dependence through dependence of the EoM on $F^2$ in (\ref{actfs}), and an implicit dependence arising from dependence of the background functions on $B$. As shown in  \cite{Gursoy:2016ofp}  the latter dependence behaves similar to the sea quarks, and the former behaves like the valence quarks.  This can be shown by isolating either of the two dependences by playing with the values of $B$ and $x$.

%%%%%%%%%%%%%%%%%%%%%%%%%%
%%%%%%%%%%%%%%%%%%%%%%%%%%
\begin{figure}
\centering
\resizebox{.45\textwidth}{!}{
 \includegraphics{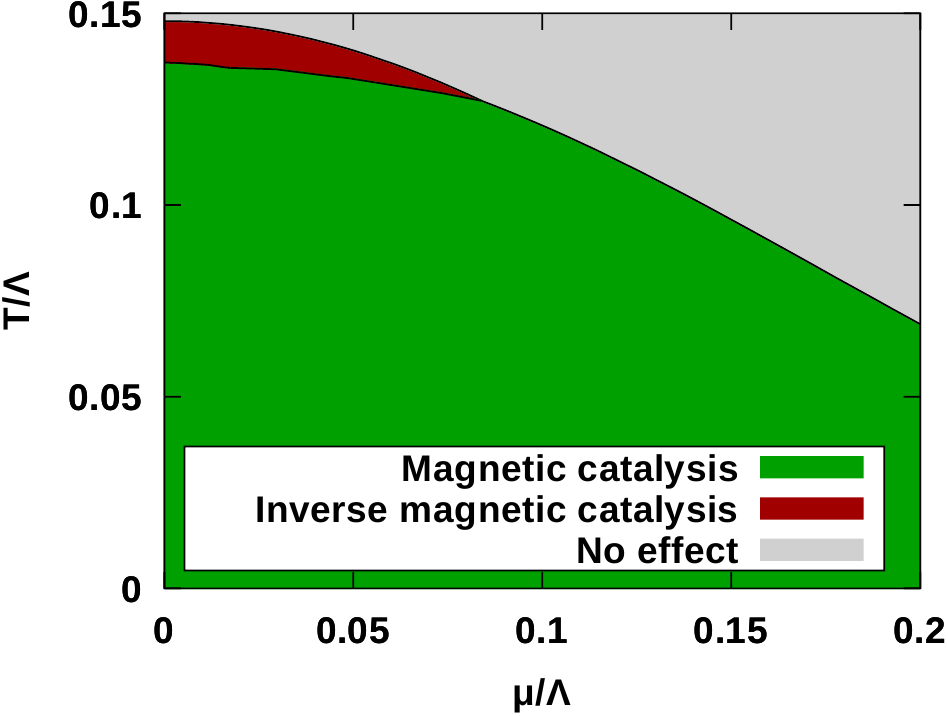}}
\vspace{.3cm}       
 \caption{Presence of inverse magnetic catalysis in improved holographic QCD at finite quark chemical potential. Figure from \cite{Gursoy:2017wzz}.}
\label{fig::8}    
\end{figure}
%%%%%%%%%%%%%%%%%%%%%%%%%%
%%%%%%%%%%%%%%%%%%%%%%%%%%

In passing, we note that the same holographic model also suggests another possible source for the phenomenon: anisotropy in the quantum state. In \cite{Giataganas:2017koz,Gursoy:2018ydr} it was shown that the same inverse catalysis effect can be reproduced in an anisotropic state in the absence of $B$. This effect, coined ``inverse anisotropic catalysis'' in \cite{Gursoy:2018ydr} hints at the possibility that inverse magnetic catalysis may be mainly due to the anisotropy created by B, rather than its impact on charge dynamics. It would be very interesting to investigate whether inverse anisotropic catalysis is present on the lattice.  

Next, one can ask what happens to inverse magnetic catalysis at finite density? This question cannot be directly addressed on the lattice because of the sign problem \cite{deForcrand:2010ys}. The question was investigated in holography in \cite{Gursoy:2017wzz} by turning on both chemical potential and magnetic field as in (\ref{VmuB}). Our findings are summarized in Figs. \ref{fig::7} and \ref{fig::8}. Figure \ref{fig::7} shows that the chiral restoration temperature decreases (increases) for small (large) $\mu$ signaling IMC for small densities. The regime where IMC is realized is shown by the red area in \ref{fig::8}.  

%%%%%%%%%%%%%%%%%%%%%%%%%%%%%%%%%%%%%%%%%%%
%%%%%%%%%%%%%%%%%%%%%%%%%%%%%%%%%%%%%%%%%%%
%%%%%%%%%%%%%%%%%%%%%%%%%%%%%%%%%%%%%%%%%%%
\section{Quark-antiquark potential}
\lab{sec::7}
%2p
%%%%%%%%%%%%%%%%%%%%%%%%%%%%%%%%%%%%%%%%%%%
%%%%%%%%%%%%%%%%%%%%%%%%%%%%%%%%%%%%%%%%%%%
%%%%%%%%%%%%%%%%%%%%%%%%%%%%%%%%%%%%%%%%%%%v

%%%%%%%%%%%%%%%%%%%%%%%%%%
%%%%%%%%%%%%%%%%%%%%%%%%%%
\begin{figure}
\centering
\resizebox{.45\textwidth}{!}{
 \includegraphics{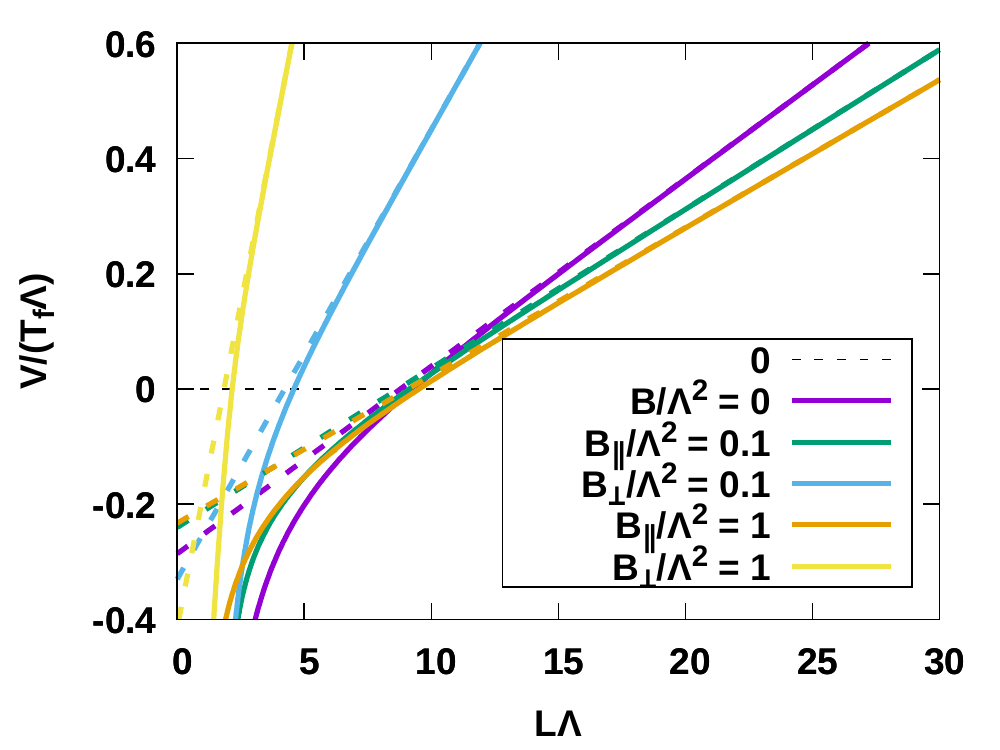}}
\vspace{.3cm}       
\caption{Quark-antiquark potentials for two cases, when the $q\bar{q}$ pair is separated along the magnetic field ($B_\parallel$) or orthogonal to it ($B_\perp$). The potentials are shown together with the asymptotic behavior of the potentials for large separation (dashed)\@. We also plot the dependence on the parameter $c$ introduced in (\ref{wl}). Figure from \cite{Gursoy:2020kjd}.}
\label{fig::9}    
\end{figure}
%%%%%%%%%%%%%%%%%%%%%%%%%%
%%%%%%%%%%%%%%%%%%%%%%%%%%

External magnetic fields also impact another important observable in QCD, the quark-antiquark potential.
In the confining ground state of QCD, quarks experience a linear potential of the form
%%%%%%%% 
\be\lab{Vqq}
V_{q\bar q}(L)  \approx  \sigma_0 L - \frac{\alpha_{eff}}{L}\, ,
\ee
%%%%%%%%                                                                    
where $\sigma_0$ is called the string tension and $\alpha_{eff}$ is an effective QCD coupling. The second term in (\ref{Vqq}) is the analog of  Coulomb interaction between particles of two opposite charges, while the first term can be understood as arising from a gluon flux tube stretching like a string between the quark and the anti-quark. This observable in the presence of a magnetic field has been studied on the lattice \cite{Bonati:2014ksa}, where it was found that $V_{q\bar q}$ becomes anisotropic. In particular, the string tension increases in the direction perpendicular to B and decreases parallel to it. 

Quark-anti-quark potential in the holographic dual description is given by the area of a Nambu-Goto test string embedded in the 5D bulk with end-points anchored to the quark and the anti-quark locations on the boundary gauge theory\footnote{This area typically diverges and should be regulated by subtracting disconnected strings attached to the quark and the anti-quark.} \cite{Maldacena:1998im,Rey:1998ik}. It was computed in  the holographic dual of ${\cal N}=4$ super Yang-Mills in the presence of magnetic field in \cite{Rougemont:2014efa}; see also \cite{Giataganas:2012zy,Giataganas:2013zaa} . 

The problem was studied in the Veneziano limit of improved holographic models in \cite{Gursoy:2020kjd}, results of which we present below. From a qualitative point of view, one first wonders if magnetic fields obstruct linear confinement, as one might think because they generically induce an AdS$_3 \times \mathbb{R}^2$ region \cite{Gursoy:2020kjd} in the geometry. This would imply a conformal rather than confining quark-anti-quark potential. This does not happen however. In contrast to Einstein-Maxwell type models~\cite{DHoker:2012rlj,Rougemont:2014efa}, in our setup $B$ enters in the 5D theory through the flavor DBI action. Since the flavor sector decouples in the deep IR, the effect of $B$ on the geometry becomes negligible, and far IR geometry is the same, that is confining, as in the case $B = 0$\@. 

%%%%%%%%%%%%%%%%%%%%%%%%%%
%%%%%%%%%%%%%%%%%%%%%%%%%%
\begin{figure}
\centering
\resizebox{.45\textwidth}{!}{
 \includegraphics{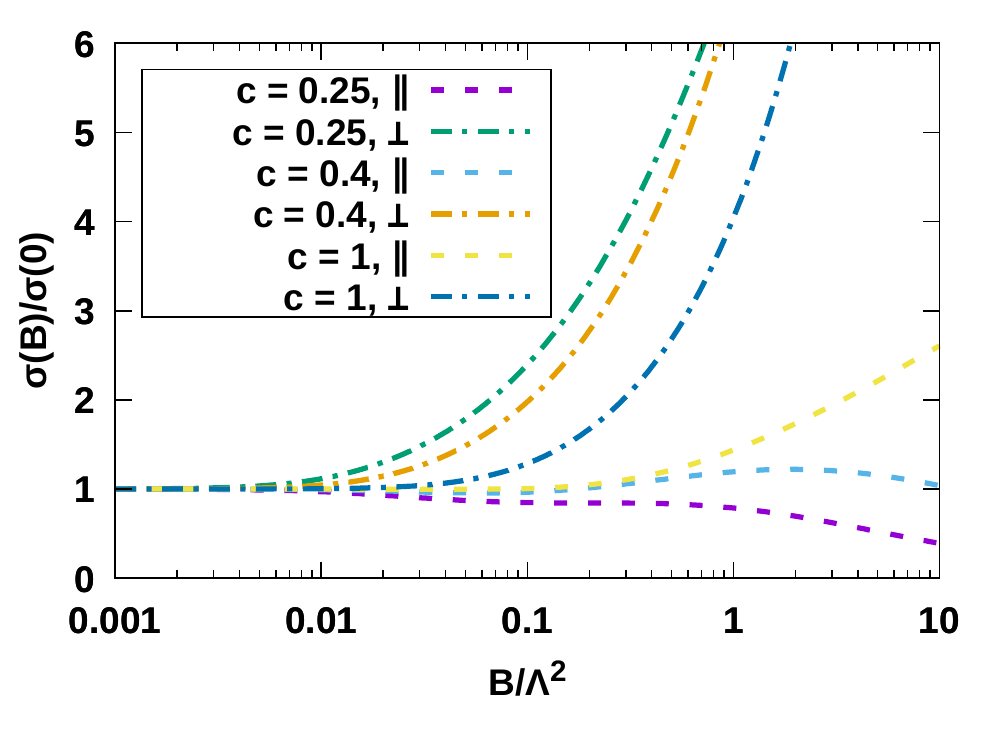}}
\vspace{.3cm}       
\caption{String tensions, i.e. the slope of the quark-anti-quark potentials in the confining regime, for the cases explained in Fig. \ref{fig::9}. Figure from \cite{Gursoy:2020kjd}.}
\label{fig::10}    
\end{figure}
%%%%%%%%%%%%%%%%%%%%%%%%%%
%%%%%%%%%%%%%%%%%%%%%%%%%%

We present the quark potential and the string tension as a function of quark-anti-quark separation $L$ and magnetic field in Figs.~\ref{fig::9} and \ref{fig::10} for $B$ parallel and perpendicular to the quark separation vector denoted by $B_\parallel$ and $B_\perp$. It is also interesting to compare the string tensions with the lattice study of \cite{Bonati:2014ksa}. In fig.~\ref{fig::10} we plot the string tensions (normalized by the $B=0$ value) for different choices of the parameter $c$ and observe that the choice $c=0.25$ agrees better with \cite{Bonati:2014ksa}. In particular, this lattice study also finds monotonically increasing (decreasing) functions of $B$ for the perpendicular (parallel) cases in the range $eB\sim0-1.2$ GeV\@. 

%%%%%%%%%%%%%%%%%%%%%%%%%%%%%%%%%%%%%%%%%%%
%%%%%%%%%%%%%%%%%%%%%%%%%%%%%%%%%%%%%%%%%%%
%%%%%%%%%%%%%%%%%%%%%%%%%%%%%%%%%%%%%%%%%%%
\section{Shear viscosity}
\lab{sec::8}
%2p
%%%%%%%%%%%%%%%%%%%%%%%%%%%%%%%%%%%%%%%%%%%
%%%%%%%%%%%%%%%%%%%%%%%%%%%%%%%%%%%%%%%%%%%
%%%%%%%%%%%%%%%%%%%%%%%%%%%%%%%%%%%%%%%%%%%

It is well known that shear viscosity to entropy density takes the universal value $\eta/s= 1/4\pi$ \cite{Policastro:2001yc,Kovtun:2003wp} in two-derivative holography. This result is in remarkable agreement with experimental data \cite{CasalderreySolana:2011us,Nijs:2020roc,Nijs:2020ors,Everett:2020xug}. This finding, among others, has been an important driving force behind the applications of holography to QGP physics. It is important to stress that this result is universal \cite{Buchel:2003tz}  in an isotropic state i.e. it is completely independent of the details of the 5D bulk action, as long as higher derivative terms can be ignored\footnote{Terms with more than two derivatives typically arise in effective actions obtained from string theory and are associated with higher powers of the string length scale in AdS units $\l_s/\ell$. They correspond to $1/\lambda$ corrections, see equation (\ref{dic1}).}. For example, $\eta/s$ will be $1/4\pi$ regardless of the choice of potentials $V$, $V_{f0}$, $\kappa$, $w$ and $a$ entering the action (\ref{fullaction}).

Yet, this is only true in an isotropic state. In an an-isotropic situation, caused, for example, by presence of an external magnetic field or different pressure gradients in different directions---which occurs in heavy ion collisions---the shear viscosity on the (xy), (xz) and (yz) planes could be all different\cite{Burikham:2016roo,Hartnoll:2016tri,Alberte:2016xja,Ling:2016ien,Baggioli:2020ljz,Erdmenger:2010xm,Rebhan:2011vd,Mamo:2012sy,Jain:2014vka,Critelli:2014kra,Jain:2015txa,Finazzo:2016mhm,Giataganas:2017koz}. In case of partial breaking of isotropy, $SO(3)\to SO(2)$, the shear viscosity/entropy ratio on the plane perpendicular to the anisotropy vector continues to assume its universal value $1/4\pi$ as the conditions for universality still apply. This, for example, yields $\eta_{xy}/s = 1/4\pi$ when $B$ is in the z-direction. In the general case, the shear viscosity tensor is computed using the Kubo formula,
%%%%%%%%
\be
\eta_{ij}=-\frac{1}{\omega}\textrm{Im}\, \langle T_{ij}(\omega,\vec{k}_1)T_{ij}(\omega,\vec{k}_2)\rangle\big|_{\omega\to0,\,\vec{k}_{1,2}\to0}\, ,
\ee
%%%%%%%%
where the limit on the right is taken first. These two point functions are, in turn, computed in holography from fluctuations of the associated metric components as explained in section \ref{sec::2}. For the metric  (\ref{geo1}) the result is (see \cite{Gursoy:2020kjd} for a recent derivation)
 %%%%%%%%%%
\be\label{etas}
\frac{\eta_{xy}}{s}=\frac{1}{4\pi}\,, \qquad
\frac{\eta_{xz}}{s}=\frac{\eta_{yz}}{s}= \frac{e^{2W(r_h)}}{4\pi}\, \qquad
\ee
 %%%%%%%%%%

%%%%%%%%%%%%%%%%%%%%%%%%%%
%%%%%%%%%%%%%%%%%%%%%%%%%%
\begin{figure}
\centering
\resizebox{.45\textwidth}{!}{
 \includegraphics{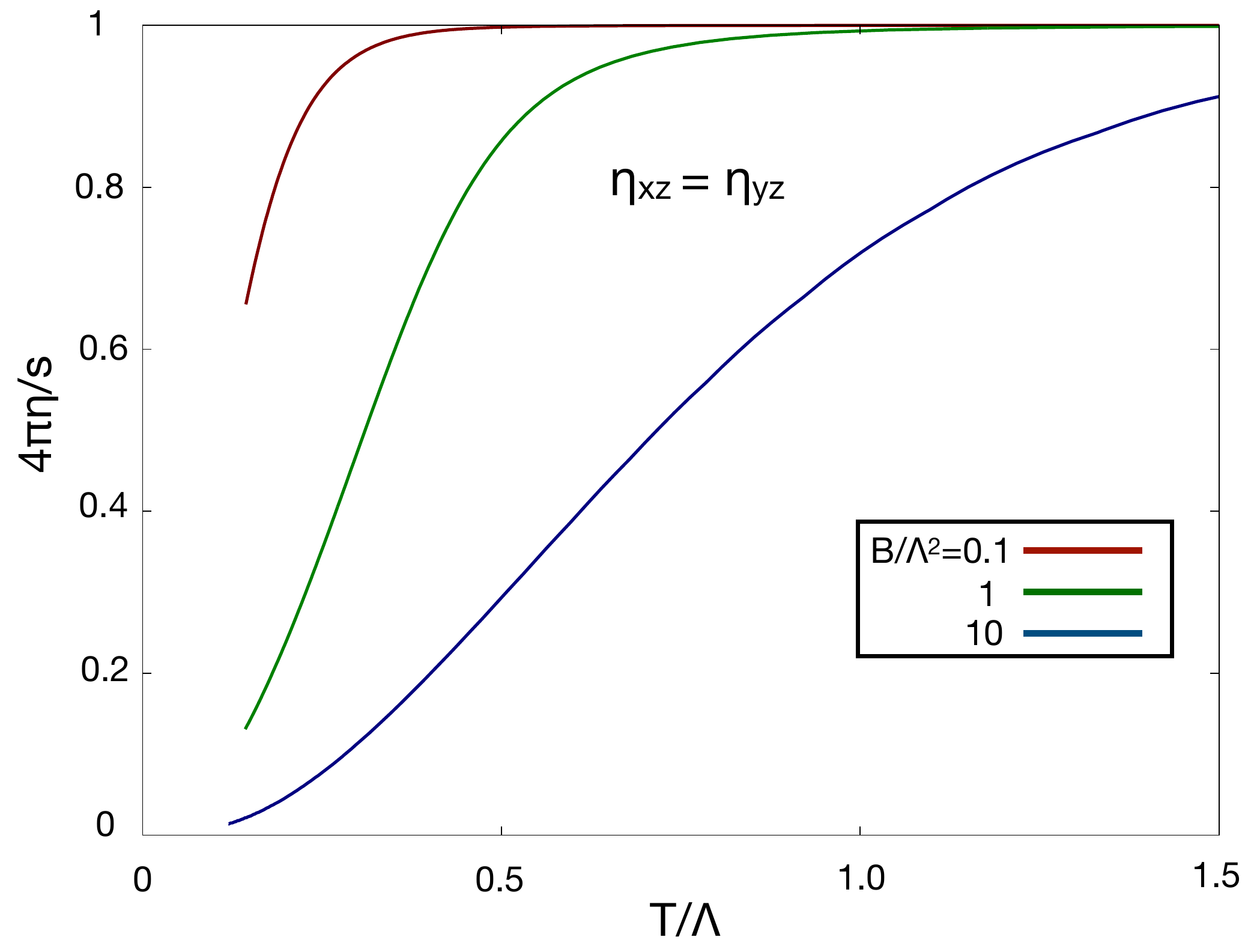}}     
\caption{Shear viscosity to entropy ratio of the longitudinal component $\eta_{xz} = \eta_{yz}$ as a function of temperature (in units of $\Lambda\sim1$ GeV) for $c=0.25$\@ in holographic QCD with a magnetic field along $z$. The curves are cut off at the chiral transition temperature below which there is a non-trivial chiral condensate. Figure from \cite{Gursoy:2020kjd}.}
\label{fig::11}    
\end{figure}
%%%%%%%%%%%%%%%%%%%%%%%%%%
%%%%%%%%%%%%%%%%%%%%%%%%%%
We plot them in Fig.~\ref{fig::11}.  The longitudinal components decrease monotonically from the UV to the IR as also observed in other anisotropic backgrounds \cite{Erdmenger:2010xm,Rebhan:2011vd,Mamo:2012sy,Jain:2014vka,Jain:2015txa,Giataganas:2017koz}.  The universal value $\eta_{ij}/s=1/4\pi$ is attained in the UV. This is because the 5D background, (\ref{geo1}), is chosen to be asymptotically AdS\@. They attain smaller, non-zero values in the IR. 
See \cite{Gursoy:2020kjd} for more on interpretation of these results. In summary, holographic calculations indicate that magnetic field substantially reduces the shear viscosity parallel to it, a fact which could have important implications for the quark-gluon plasma created in heavy ion collisions.

%%%%%%%%%%%%%%%%%%%%%%%%%%%%%%%%%%%%%%%%%%%
%%%%%%%%%%%%%%%%%%%%%%%%%%%%%%%%%%%%%%%%%%%
%%%%%%%%%%%%%%%%%%%%%%%%%%%%%%%%%%%%%%%%%%%
\section{Discussion}
\lab{sec::9}
%1p
%%%%%%%%%%%%%%%%%%%%%%%%%%%%%%%%%%%%%%%%%%%
%%%%%%%%%%%%%%%%%%%%%%%%%%%%%%%%%%%%%%%%%%%
%%%%%%%%%%%%%%%%%%%%%%%%%%%%%%%%%%%%%%%%%%%

We reviewed recent progress in the holographic approach to strong nuclear force in the presence of magnetic fields, mainly focusing on thermodynamic and transport properties relevant to the quark-gluon plasma. We discussed the phase diagram of large-N QCD, magnetic catalysis of chiral condensate, quark-anti-quark potential and shear viscosity in the presence of external magnetic fields. As the first-principles lattice theory is already available to study QCD with magnetic fields, our results are mainly useful in regimes where the lattice approach becomes unsuitable, i.e. finite quark density, transport and large magnetic fields. 

The premier promise of holography is to uncover universal phenomena at strong coupling, and explain these phenomena in a qualitative manner. Several such examples are listed at the end of section \ref{sec::3}. We should ask whether there are similar examples with magnetic fields. The answer is in the affirmative. One such example is inverse magnetic catalysis. As discussed in detail in section \ref{sec::6}, improved holographic models generically exhibit this phenomenon for sufficiently small values of the parameter $c$ in the $w$ potential of the DBI action. Interestingly, for larger values of $c$ for which IMC disappears, the quark-anti-quark potential profile also becomes unphysical. This is an example of how holography can relate different phenomena in the same theory by requiring overall consistency. The model also points toward a possible explanation of the IMC phenomenon by relating it to backreaction of B on the background geometry, which was suggested in \cite{Gursoy:2016ofp} as the holographic analog of the ``sea quarks'' \cite{Bruckmann:2013oba,Bruckmann:2013ufa}. We did not cover here another universal feature of the holographic model, that the chiral condensate tends to decrease in an anisotropic state even in the absence of B, called ``inverse anisotropic catalysis'' \cite{Giataganas:2017koz,Gursoy:2018ydr}.  

Another generic behavior we observe is the presence of a deconfined but chiral symmetry breaking state in the phase diagram, shown by blue in Figs. \ref{fig::2}, \ref{fig::3} and \ref{fig::n1}. Fig. \ref{fig::2} shows that this phase reappears for large values of $B$, and, for very large $B$, it exhibits magnetic catalysis rather than the inverse effect. This is consistent with the QCD result obtained by solving the gap equation in the ``improved rainbow approximation'', which is valid for $eB \gg \Lambda_{QCD}^2$, see e.g.  \cite{Miransky:2015ava}. Finally, we observe in section \ref{sec::8} that the shear viscosity on the plane parallel to B generically decreases, and it does so substantially. This observation may be relevant for off-central heavy ion collisions. 

Beyond qualitative insights and a generic, broad-brush picture of observables, improved holographic models also seem to provide accurate quantitative results, at least for the thermodynamic potentials, see Fig. \ref{fig::n3}. This can hardly be a coincidence. After all, it is inconceivable that QCD in full --- with infinitely many operators mixing in the RG flow --- could be described by gravity coupled to a few scalars.  Why does it, then, seem to work? Holography in the two-derivative approximation certainly cannot describe the entire energy range of QCD. The perturbative UV regime requires string corrections as clear from the fact that the two-derivative answer  for the shear viscosity to entropy ratio, $1/4\pi$, is largely off the pQCD answer. However, there are strong arguments --- mostly due to Polyakov, see e.g. \cite{Polyakov:1998ju,Polyakov:2001af} --- that large N QCD is dual to 5D non-critical string theory (possibly with world-sheet supersymmetry) which might be approximated in the IR, reliably, by Einstein's gravity coupled to several scalars and form-fields. A hint in QFT is that, Ward identities select a  closed subset of operators, e.g. stress tensor and $\tr G^2$ in pure Yang-Mills. More generally, QCD sum-rules \cite{Shifman:1978bx} indicate a semi-closed subset of operators dominating the IR regime. Thus, it is sensible that, a subset of UV insensitive observables are well captured by only a few relevant or marginal operators in the background of infinitely many other, whose overall effect in the gravity dual is to dress the potentials entering the two-derivative action. Thermodynamic potentials turn out to be this type. We refer to \cite{Kiritsis:2009hu} for a detailed discussion.   

A shorter argument is the following. If the IR dynamics in the plasma state is governed by relativistic hydrodynamics, which is, by definition, characterized by the ``slow'', IR variables, then one can construct,  by hand, a corresponding 5D gravitational theory following the fluid-gravity prescription \cite{Bhattacharyya:2008jc}. Improved holography does not use this prescription but plausibly arrives at the same answer. 

We omitted a number of important topics in this review. Non-equilibrium dynamics of strongly interacting plasma coupled to electromagnetic fields \cite{Fuini:2015hba,Endrodi:2018ikq} is a major problem with a bearing on both heavy ion collisions and BNS mergers.  AdS/CFT generically predicts rapid ``hydrodynamization'' of the debris left by the colliding heavy ions, in a time scale $\tau\sim 1/T$ \cite{Chesler:2008hg,Heller:2011ju} and a natural question is whether $B$ affects the proportionality coefficient markedly or not. This is an open problem. 

Another related question involves the significance of anomalous transport in quark-gluon plasma, e.g. chiral separation, chiral magnetic and chiral vortical effects, and chiral magnetic wave, see \cite{Kharzeev:2015znc,Kharzeev:2020jxw} for a review. Consider chiral magnetic effect for definiteness. The magnitude of the chiral magnetic current $J = a_{CME}\, B$ depends on the value of the coefficient $a_{CME}$, which, if electromagnetic fields are treated in the linear approximation, mainly depends on the chiral imbalance in the system. This is, in turn, to a great extent determined by the sphaleron decay rate, that was computed at weak coupling \cite{Bodeker:1999gx,Bodeker:1998hm,Bodeker:1999gx}, and in strongly coupled conformal ${\cal N}=4$ super Yang-Mills plasma without, and with magnetic fields \cite{Son:2002sd,Basar:2012gh} using holography. The calculation was carried out in improved holographic models again without and with magnetic fields \cite{Gursoy:2012bt,Drwenski:2015sha} where it was shown that the decay rate goes up significantly both as a result of non-conformality and of magnetic fields. Eventually, the amount of chiral imbalance in the system, hence the stre-ngth of anomalous transport effects should be determined by the pre-equilibrium physics. This is another open problem. 

Magnetic fields also influence other interesting observables such as the entanglement entropy and the butterfly velocity that can be studied using holography. In \cite{Gursoy:2020kjd} these observables were proposed as tools to disentangle the effect of pressure anisotropy and magnetic fields in heavy ion collisions. Finally, the techniques we discussed in this review have interesting applications in condensed matter \cite{Hartnoll:2009sz,Zaanen:2015oix} where magnetic fields provide crucial probes, for example, of quantum phase transitions. 

We conclude this review with a look forward.  Recently opened fascinating windows into strong nuclear matter, LIGO/Virgo GW detectors, NICER measurements, ongoing (recently upgraded LHC, RHIC) and future planned large-scale heavy ion experiments (FAIR, NICA), all probe, directly or indirectly, its electromagnetic properties, making today an exciting time to construct the theory. We strongly believe a judicious combination of lattice QCD, kinetic theory, hydrodynamics with holography will deliver the basic observables in these measurements: particle yields in heavy ion collisions and gravitational waveforms in neutron star mergers.

\section*{Acknowledgements}

This review grew out of a long series of work  in collaboration mainly with Elias Kiritsis, Francesco Nitti, Liuba Mazzanti and Matti Jarvinen but also including Tuna Demircik, Tara Drwenski, Ioannis Iatrakis, Aron Jansen, Georgios Michalogiorgakis, Govert Nijs, Andy O'Bannon, Marco Panero, Raimond Snellings, Andreas Schafer and Wilke van der Schee to all of whom I am grateful. The author is partially supported by the Delta-Institute for Theoretical Physics (D-ITP), both funded by the Dutch Ministry of Education, Culture and Science (OCW).

\appendix

%%%%%%%%%%%%%%%%%%%%%%%%%%%%%%%%%%%%%%%%%%%
%%%%%%%%%%%%%%%%%%%%%%%%%%%%%%%%%%%%%%%%%%%
%%%%%%%%%%%%%%%%%%%%%%%%%%%%%%%%%%%%%%%%%%%
\section{The Potentials}
\label{app::1}%
%%%%%%%%%%%%%%%%%%%%%%%%%%%%%%%%%%%%%%%%%%%
%%%%%%%%%%%%%%%%%%%%%%%%%%%%%%%%%%%%%%%%%%%
%%%%%%%%%%%%%%%%%%%%%%%%%%%%%%%%%%%%%%%%%%%

In this appendix we list the potentials of the holographic model. Defining $\lambda = \exp \f$, the potentials are: 

\begin{eqnarray}
\label{Vf0SB}
V(\lambda)&=&{12\over \ell^2}\biggl[1+{88\lambda\over27}+{4619\lambda^2
\over 729}{\sqrt{1+\ln(1+\lambda)}\over(1+\lambda)^{2/3}}\biggr]\, , \\
 V_{f0}(\lambda)& =& {12\over \mathcal{L}_{UV}^2}\biggl[{\mathcal{L}_{UV}^2\over\ell^2}
-1+{8\over27}\biggl(11{\mathcal{L}_{UV}^2\over\ell^2}-11+2x \biggr)\lambda\nn\\
 &&+{1\over729}\biggl(4619{\mathcal{L}_{UV}^2\over \ell^2}-4619+1714x - 92x^2\biggr)\lambda^2\biggr] \, , \nn \\
 \kappa(\lambda) &=& {[1+\ln(1+\lambda)]^{-1/2}\over[1+\frac{3}{4}(\frac{115-16x }{27}-{1\over 2})\lambda]^{4/3}} \,, \ a(\lambda)=\frac{3}{2 \, \mathcal{L}_{UV}^2} \, ,
\label{kappaa}
 \end{eqnarray}
where $\mathcal{L}_{UV}$ is given in terms of the AdS radius $\ell$, such that the boundary expansion of the metric is 
$ A \sim \ln \left( { \mathcal{L}_{UV} / r} \right)+\cdots \, .$
The radius depends on $x$ as 
\be
\mathcal{L}_{UV}^3 = \ell^3 \left( 1+{7 x \over 4} \right) \, .
\label{adsrad}
\ee
The function $w$ is parametrized by a single parameter $c$ 
\be
w(\lambda)=\kappa(c\lambda) =  \frac{( 1+\log(1+ c \, \lambda))^{-{1\over 2}}}{\left(1+ {3 \over 4} \left({115-16 x \over 27}-{1\over 2} \right) c \,\lambda \right)^{4/3}}       \, ,
\label{wl}
\ee
where $x$ is the ratio of the number of flavors to color. 
%\end{appendix


%\begin{figure}
%\resizebox{0.75\textwidth}{!}{%
%  \includegraphics{leer.eps}
%\vspace{5cm}       % Give the correct figure height in cm
%\end{figure}

%\begin{figure*}
%\vspace*{5cm}       % Give the correct figure height in cm
%\end{figure*}
%

% BibTeX users please use
 \bibliographystyle{epj}
 \bibliography{magrev}
%\begin{thebibliography}{}
%Author, \textit{Book title} (Publisher, place year) page numbers
% etc
%\end{thebibliography}

\end{document}